\documentclass[a4paper,11pt]{article}
\usepackage{jheppub} 
\usepackage{lineno}
\usepackage{titlesec}
\usepackage{tabularx}
\usepackage{pdfpages}
\usepackage{float}
\usepackage{subcaption}
\usepackage{multirow}
\usepackage{booktabs}
\usepackage{makecell}
\usepackage[numbers,sort&compress]{natbib}  
\usepackage{todonotes}

\usepackage[english]{babel}

\newcommand{\nn}{\nonumber}
\newcommand{\dm}{d}
\newcommand{\nl}{l}
\newcommand{\nh}{h}

\preprint{USTC-ICTS/PCFT-25-52}
\title{\boldmath Application of Transformer in 2D lattice Yang-Mills theory}

\author[a,b,c]{Zeyu Li,}

\author[b,c]{Guorui Zhu,}

\author[j]{Wenjie He,}

\author[d,e,f,g]{Bo Feng,}

\author[h]{Jiaqi Chen,}

\author[f]{Ming-xing Luo,}

\author[b,c,g,i]{Gang Yang}

\affiliation[a]{Center for High Energy Physics, Peking University, Beijing 100871, China}
\affiliation[b]{Institute of Theoretical Physics, Chinese Academy of Sciences, Beijing 100190, China}
\affiliation[c]{School of Physical Sciences, University of Chinese Academy of Sciences, Beijing 100049, China}
\affiliation[d]{State Key Laboratory of Nuclear Physics and Technology, Institute of Quantum Matter, South China Normal University, Guangzhou 510006, China}
\affiliation[e]{Guangdong Basic Research Center of Excellence for Structure and Fundamental Interactions of Matter, Guangdong Provincial Key Laboratory of Nuclear Science, Guangzhou 510006, China}
\affiliation[f]{Beijing Computational Science Research Center, Beijing 100193, China}
\affiliation[g]{Peng Huanwu Center for Fundamental Theory, Hefei, Anhui, 230026, China}
\affiliation[h]{Beĳing Key Laboratory of Optical Detection Technology for Oil and Gas, China University of PetroleumBeĳing, Beĳing 102249, China}
\affiliation[i]{School of Fundamental Physics and Mathematical Sciences, Hangzhou Institute for Advanced Study, UCAS, Hangzhou 310024, China}
\affiliation[j]{Department of Physics, Beijing Normal University, Beijing 100875, China}

\abstract{
	We employ the Transformer to learn patterns in two-dimensional lattice Yang-Mills theory. Specifically, we represent both Wilson loops and their expectation values as tokenized sequences. Taking the shape of Wilson loops as input, the model successfully predicts expectation values with high accuracy, demonstrating that the Transformer can effectively learn the supervised mapping from loop geometry to analytic results within the training distribution. We note that the 2D theory is exactly solvable and the target expressions belong to a restricted set of polynomial forms, which considerably simplifies the learning task. Our study differs from prior machine learning applications in lattice QCD by emphasizing analytical structures rather than numerical computations. We explore model performance under varying hyperparameters, training data sizes, and sequence lengths. This work serves as a proof-of-concept study toward extending such methods to higher dimensions and inspiring rigorous analytical derivations.
}

\begin{document}
	\maketitle
	\setcounter{footnote}{0}
	
\section{Introduction}
Machine learning has long been widely applied in scientific research and has yielded numerous significant achievements. In recent years, high-energy physics has also benefited from the development of machine learning, not only in the analysis of collision events \cite{Butter:2022rso}, but also in attempts to apply it directly to high-energy physics calculations, such as differential equations of Feynman integral \cite{Calisto:2023vmm}, Feynman diagram as computational graphs \cite{Hou:2024vtx,Mitchell:2022yxp}, refining seeds of reducing Feynman integrals \cite{vonHippel:2025okr, Song:2025pwy}, lattice QCD sampling \cite{Favoni:2022Lattice, Abbott:2024kfc}, and so on. One of the uses of machine learning in scientific research is to assist in the search for laws and patterns. A successful case of such application is the study of "murmurations" in elliptic curves \cite{He_2022_1,He_2022_2,He_2023,He_2024,zubrilina2023murmurations,He_2022_3}.
In this case, AI-assisted pattern discovery can be divided into three steps:
Step one, train a machine learning model to see if it can predict the target variable from the given features \cite{He_2022_3,He_2022_5}. If successful, it indicates that the information about the target variable is embedded within the given features. Then, we can proceed to step two: further analyze the relationship between the provided features and the target variable. This step can also leverage machine learning tools—for example, Principal Component Analysis (PCA) was employed in \cite{He:2022pqn};
Step 3 uses the results from the first two steps as clues to inspire rigorous derivation of new patterns, \cite{zubrilina2023murmurations} as an example. To derive patterns with these clues is usually significantly easier than without them.
In this paper, we follow these steps and apply machine learning to study the properties of lattice Yang-Mills theory, focusing on the first step as a proof of concept.

Non-perturbative QCD has always been an important research area in theoretical physics, as it holds significant implications for understanding the phenomena of color confinement and the mass gap problem. Over the past few decades, lattice QCD has been the most important strategy for studying non-perturbative QCD. By discretizing spacetime, we can numerically compute many physical quantities non-perturbatively using Monte Carlo methods, such as the hadron mass spectrum \cite{PhysRevD.74.014504,PhysRevD.79.054501,Fodor:2012gf}, the anomalous magnetic moment of the muon \cite{Procura:2019cdx,Borsanyi_2021}. Nevertheless, analytical calculations in lattice QCD remain extremely challenging and require further research and understanding. 
Currently, analytical solutions are only possible for some simple toy models such as two-dimensional lattice Yang-Mills theory \cite{PhysRevD.21.446,Friedan:1980tu}. 
However, in the meantime, some semi-analytical methods such as the bootstrap approach have been developed \cite{Anderson_2017, Kazakov:2022xuh}. 
Based on positivity, this method applies to arbitrary 
dimensions and demonstrates good convergence in certain theories. For example, U$(1)$, SU$(2)$ and SU$(3)$ lattice gauge theory has been well studied by bootstrap method \cite{Kazakov:2024ool,Guo:2025fii,Li:2024wrd}.

In lattice Yang-Mills theory, we need to compute the expectation value of Wilson loops, which are the products of link variables.
A natural question is whether the expectation value bears some simple relationship to the shape of the Wilson loop, so that one could directly derive the desired outcome from the shape alone.
While the methods above can calculate the corresponding value of the Wilson loop, the relationship between its shape and the computed value remains unclear. Even in two dimensions, the analytical results can, in principle, be derived from the shape; however, the intricate intermediate steps leave the relationship ambiguous. In other words, the result is not directly expressed as a function of shape-related parameters such as area or overlapping area. In higher dimensions, the situation becomes even more complex and is not yet well understood.
In this article, we aim to explore this problem. However, the scope of this work is not so ambitious as to directly solve the problem. Instead, we first attempt to use machine learning in two dimensions to learn the mapping from shape to expectation value, training a model capable of predicting the result directly from shape information. We should note that the 2D theory is exactly solvable, and as we will discuss in section \ref{sec:discussion}, the analytic expressions for Wilson loops up to length 16 belong to a limited set of polynomial forms, which significantly constrains the complexity of the learning task. Nevertheless, we hope the result could be a useful proof-of-concept step toward the following two goals. First, we hope the approach could be extended to higher dimensions in the future; Second, it could provide clues for deriving a more rigorous relation between the expectation value and the shape information.

Among the machine learning strategies, we chose the architecture called ``Transformer''. Transformer \cite{vaswani2023attentionneed} is a deep learning architecture based on the self-attention mechanism. It relies entirely on attention mechanisms to capture global dependencies in input sequences, significantly improving the efficiency and performance of long-sequence modeling. Today, transformers have not only become a cornerstone in natural language processing (NLP) but have also been widely applied to cross-modal tasks such as computer vision and speech recognition. Its powerful capabilities have also attracted scholars in high-energy physics to apply it in this field.
Researchers have explored its applications in computing the modulus square of scattering amplitudes \cite{Alnuqaydan_2023}, simplifying polylogarithms \cite{dersy2022simplifyingpolylogarithmsmachinelearning}, learning the structure of symbols in high-loop scattering amplitudes \cite{Cai:2024znx}, and simplifying the helicity-amplitude \cite{Cheung_2025}. 
In all these cases, it has demonstrated remarkably high accuracy. In our case, the shape information of Wilson loops could naturally be represented as tokens in a transformer model, or more precisely, as an ordered sequence of symbols (we will present the chosen representation later). Therefore, we adopt this framework to learn the mapping from Wilson loop shapes to their analytic expectation values in 2-dimensional lattice Yang-Mills theory.
To be more specific, we train the transformer on the shape of the Wilson loop as a feature and their corresponding analytical results as the target variable. After training, we observed that the model could predict the expectation values of Wilson loops in the test set with high accuracy. 
This high accuracy demonstrates that the Transformer can effectively reproduce the known analytic results from shape information alone. As we discuss later, the model can generalize to geometrically distinct Wilson loops that share the same analytic structure, though it cannot extrapolate to entirely new polynomial forms beyond its training data. 
We also mention that there have been some studies applying machine learning to lattice QCD, such as Monte Carlo sampling \cite{albergo2021introductionnormalizingflowslattice} and the integration of generative models with stochastic quantization \cite{Zhu:2024kiu,Aarts:2024agm,Zhu:2025pmw,Wang:2023exq}. Our research distinguishes itself by focusing on the analytical structures of lattice theory.

The structure of this paper is organized as follows. In section \ref{sec:physical_setup}, we will briefly review the basic concept of lattice field theory and introduce the structure of two-dimensional lattice Yang-Mills theory. In section \ref{sec:conventions_and_setup}, we will introduce the strategy of tokenizing Wilson loops and their corresponding analytic solution. In section \ref{sec:overview_of_learning_result}, we will give an overview of the learning result. In section \ref{sec:hyperopt}, we will discuss how hyperparameters influence the learning efficiency. In section \ref{sec:fractrain}, we train models with varying proportions of training data. In section \ref{sec:mixtrain}, we try to enhance the model's ability by mixing Wilson loops with different lengths. 
In section \ref{sec:discussion}, we give some discussion and outlooks.

\section{Physical Setup}\label{sec:physical_setup}

In this section, we briefly review the lattice gauge theory \cite{Wilson:1974sk}. We will not consider quarks and mainly focus on the result of two-dimensional lattice Yang-Mills theory, in which the Wilson loop operators will be the object for machine learning.

\subsection{Basis of lattice gauge field theory}

In lattice field theory, we discretize the space-time and study the field theory on a lattice. A $d$-dimension Euclidean lattice can be represented as:
\begin{equation}
	\Lambda_d = \{x| x = a\sum_{\mu = 1}^{d} n_{\mu}\hat{\mu} \}  \,,
\end{equation}
where $\hat{\mu}$ is the unit vector of the $\mu$th direction and the components of $n_{\mu}$ being integer. We are interested in the pure Yang-Mills theory living on this lattice, where the gluon field $A^{a,\mu}(x)$ in continuous space-time will become the ``gauge link'' $U_{\mu}(x)$ connecting two sites in the lattice. The notation $U_{\mu}(x)$ represents the link from point $x$ 
to $x + a \hat{\mu}$.
In SU(N) pure Yang-Mills theory, all the gauge links $U_{\mu}(x)$ will be SU(N) matrices. For the same link with opposite orientation, say $U_{\bar{\mu}}(x+a\hat{\mu})$, it will be the Hermitian conjugate of the original one:
\begin{equation}
	U_{\bar{\mu}}(x+a\hat{\mu}) = U^{\dagger}_{\mu}(x) = U^{-1}_{\mu}(x)  \,.
\end{equation}
The lattice constant $a$ will not be important in the following discussions, so we will set $a = 1$ in the rest of this paper.

The gauge transformation to a gauge link is:
\begin{equation}
	U_{\mu}(x) \to U^{\prime}_{\mu}(x) = \Omega(x) U_{\mu}(x) \Omega^{\dagger}(x+\hat{\mu}).
\end{equation} 
where $\Omega(x)$ is also a SU(N) matrix and can be chosen independently at each lattice site $x$. A physical operator is formed by the product of such links. In order to ensure the gauge invariance of physical operators, the links must form a closed loop, which is called the ``Wilson loop''. The simplest non-trivial Wilson loop is the ``plaquette":
\begin{equation}
	{\rm Tr}U_{P}(x) = {\rm Tr}(U_{\mu}(x)U_{\nu}(x+\hat{\mu}) U_{\bar{\mu}}(x+\hat{\mu}+\hat{\nu})U_{\bar{\nu}}(x+\hat{\nu}))  \,,
\end{equation}
where we use the subscript ``P'' to denote a plaquette.
Plaquette is also the simplest gauge invariant quantity in the lattice.
The action for the pure Yang-Mills lattice theory is:
\begin{equation}
	S = \frac{N}{2\lambda} \sum_{P}({\rm Tr}U_{P} + {\rm Tr}U^{\dagger}_{P})  \,,
\end{equation}
where $N$ is the number of color and $\lambda$ is the coupling constant. The summation is over all the plaquettes in the lattice.

In a quantized lattice Yang-Mills theory, the dynamics of the total system is governed by the partition function:
\begin{equation}
	Z = \int DU e^{-S} = \int DU e^{-\frac{N}{2\lambda} \sum_{P}({\rm Tr}U_{P} + {\rm Tr}U^{\dagger}_{P})}  \,,
\end{equation}
where $DU$ is the product of the Harr measure of each link variable. The physical quantity we want to study is the vacuum expectation value of an operator:
\begin{equation}
	\langle O \rangle = \frac{1}{Z}\int DU O e^{-\frac{N}{2\lambda} \sum_{P}({\rm Tr}U_{P} + {\rm Tr}U^{\dagger}_{P})}  \,.
\end{equation}
In particular, we will concentrate on the expectation value of Wilson loops:
\begin{equation}\label{WL_expect}
	\langle \mathcal{W_C} \rangle = \frac{1}{Z} \int DU \frac{1}{N} {\rm Tr}(U_{\mu_1} \ldots U_{\mu_L}) e^{-\frac{N}{2\lambda} \sum_{P}({\rm Tr}U_{P} + {\rm Tr}U^{\dagger}_{P})}  \,,
\end{equation}
where $\mathcal{C}$ represents a path $\mu_1 \ldots \mu_L$ which should be closed.

\subsection{2D lattice Yang-Mills theory}

In general dimension, the expectation value of Wilson loops in eq.\eqref{WL_expect} is very hard to evaluate and one can use the Monte Carlo method to calculate it numerically. But in two dimensions, the theory is significantly simplified. As we will see, in 2D lattice Yang-Mills theory, the expectation value of the Wilson loop can be factorized into a combination of contributions from the plaquettes.

In two dimensions, there only exist two different directions $\mu_0$ and $\mu_1$. It's convenient to make the gauge choice:
\begin{equation}
	U_{\mu_0}(x) \equiv 1   \label{eq:gauge1} \,,
\end{equation}
which can also be understood as $A_0 = 0$. In this gauge, the action becomes:
\begin{equation}
	S = \frac{N}{2\lambda}\sum_{x} {\rm Tr}( U_{\mu_1}(x) U^{\dagger}_{\mu_1}(x+ \hat{\mu}_0) + \text{H.c.})  \,.
\end{equation}
If we change the variables to:
\begin{equation}
	U_{\mu_1}(x+ \hat{\mu}_0) = U_{\mu_1}(x+ \hat{\mu}_0) U^{\dagger}_{\mu_1} (x) U_{\mu_1} (x) = 
	U_P(x)U_{\mu_1} (x) \,,
\end{equation}
Here, $U_p(x)$ is the plaquette starting from $x$ in the clockwise direction. We can easily find that the partition function will become:
\begin{equation}\label{partition_funtion_factor}
	Z = \int D U_P \; e^{-\frac{N}{2\lambda} \sum_{P}({\rm Tr}U_{P} + {\rm Tr}U^{\dagger}_{P})}
\end{equation}
The difference is that now the integral measure is changed into ``plaquette matrix'', which is the property of Haar measure:
\begin{equation}
	D(U) = D(UV) = D(VU)  \,,
\end{equation}
where $V$ is an arbitrary SU$(N)$ matrix.

The partition function in eq.\eqref{partition_funtion_factor} form can be factorized because the integrand is decoupled:
\begin{equation}
	Z = \prod_x \int dU_P(x)  e^{-\frac{N}{2\lambda} ({\rm Tr}U_{P} + {\rm Tr}U^{\dagger}_{P})} = (z)^{\frac{V}{a^2}}  \,,
\end{equation}
where $z=\int dU_P(x)  e^{-\frac{N}{2\lambda} ({\rm Tr}U_{P} + {\rm Tr}U^{\dagger}_{P})}$ represents partition function for a single plaquette, 
$V$ is the ``volume'' of the lattice an $\frac{V}{a^2}$ is the number of plaquettes. 
This means that the total partition function is the trivial product of the partition function within a single plaquette.

Not only the partition function, but also the Wilson loops will factorize into products of Wilson loops defined on a single plaquette. As a concrete example, we will show the factorization of the rectangle Wilson loop \cite{PhysRevD.21.446}. Let us consider the $T \times L$ rectangle Wilson loop with $T$ links in direction 0 and $L$ links in direction 1. With the gauge choice of \eqref{eq:gauge1}, it can be represented as:
\begin{align}\label{W_TL_origial}
	\mathcal{W}_{T\times L} &= \frac{1}{N} \langle {\rm Tr} U_{\mu_1}(T,0) \cdots U_{\mu_1}(T,L-1) U^{\dagger}_{\mu_1}(0,L-1) \cdots U^{\dagger}_{\mu_1}(0,0) \rangle \nn \\
	&= \frac{1}{N} \langle {\rm Tr} U_{\mu_1}(T,0) \cdots U_{\mu_1}(T,L-1)  \rangle \,,
\end{align}

where in the second equation we make a further gauge choice $U_{\mu_1}(0,i) = 1$ for simplicity\footnote{In lattice gauge theory, we can make a gauge choice that a ``maximal tree'' becomes identity \cite{Gattringer:2010zz}, and $A_0=0$ is a subset of such maximal tree.}. Furthermore, we recursively replace each $U_{\mu_1}(T,0)$ with following identity:
\begin{equation}
	U_{\mu_1}(t,l) = U_{P}(t,l) U_{\mu_1}(t-1,l)
\end{equation}
where we denote $U_{P}(t,l)$ as the plaquette starting from point $(t,l)$ with link $U_{\mu_1}(t,l)$. For example, the link $U_{\mu_1}(T,0)$ will become:
\begin{align}
	U_{\mu_1}(T,0) &= U_{P}(T,0) U_{\mu_1}(T-1,0) \\  \nn
	&= U_{P}(T,0)  U_{P}(T-1,0)  U_{\mu_1}(T-2,0)  \\ \nn
	&= U_{P}(T,0)  U_{P}(T-1,0) \cdots U_{P}(1,0) U_{\mu_1}(0,0) \\ \nn
	&= \prod_{t = T}^{1}  U_{P}(T,0)  \,,
\end{align}
where we have use the gauge choice $U_{\mu_1}(0,i) = 1$ in the last equality.

After replacing all the $U_{\mu_1}(T,l)$ in eq.\eqref{W_TL_origial}, $\mathcal{W}_{T\times L}$ becomes:
\begin{equation}\label{W_TL}
	\mathcal{W}_{T\times L} = \frac{1}{N} \langle {\rm Tr} \prod_{l=0}^{L-1}\prod_{t=T}^{1} U_{P}(t,l) \rangle  \,.
\end{equation}

To show the factorization of $\mathcal{W}_{T\times L}$, we consider the integration over a certain plaquette:
\begin{equation}
	\int d U_{P} e^{-\frac{N}{2\lambda} ({\rm Tr}U_{P} + {\rm Tr}U^{\dagger}_{P})} {\rm Tr} (A\; U_P \; B) \,,
\end{equation}
where $A$ and $B$ are products of other plaquettes. By requiring gauge invariance, we change $U_P$ into $V U_P V^{\dagger}$ and the integral is invariant:
\begin{align}
	\int d U_{P} e^{-\frac{N}{2\lambda} ({\rm Tr}U_{P} + {\rm Tr}U^{\dagger}_{P})} {\rm Tr} (A\; U_P \; B) &=  \int d (V U_{P} V^{\dagger}) e^{-\frac{N}{2\lambda} ({\rm Tr}U_{P} + {\rm Tr}U^{\dagger}_{P})} {\rm Tr} (A\; V U_P V^{\dagger}\; B) \nonumber \\
	&= \int d  U_{P}  e^{-\frac{N}{2\lambda} ({\rm Tr}U_{P} + {\rm Tr}U^{\dagger}_{P})} {\rm Tr} (A\; V U_P V^{\dagger}\; B) \nonumber \\
	&= \int dV \int d U_{P} e^{-\frac{N}{2\lambda} ({\rm tr}U_{P} + {\rm tr}U^{\dagger}_{P})} {\rm Tr} (A\; V U_P V^{\dagger}\; B)  \,,
\end{align}
where in the second equation, we use the property of Haar measure and insert $1 = \int dV$ in the third equation.

We then exchange the order of integration and use the following identity:
\begin{equation}
	\int dV V_{ij} V^{\dagger}_{kl} = \frac{1}{N} \delta_{il}\delta_{jk}  \,.
\end{equation}
After integration of V the original ${\rm Tr} (A\; U_P \; B)$ can be replaced by $\frac{1}{N} {\rm Tr}(U_P) {\rm Tr}(BA)$. In $\text{Tr}(BA)$, there is no $U_P$, so we can further consider integration over other plaquettes and repeat the above procedure. Finally, we will find that $\mathcal{W}_{T\times L}$ factorize as follows:
\begin{equation}
	\mathcal{W}_{T\times L} = \prod_{l=0}^{L-1}\prod_{t=T}^{1} \langle \frac{1}{N} {\rm Tr}  U_{P}(t,l) \rangle = \langle \frac{1}{N} {\rm Tr}  U_{P} \rangle^{T \times L}  \,,
\end{equation}
which means that the rectangular loop is just the product of the plaquettes it contains. A simple example is:
\begin{align}\label{factorize_1}
	\langle \;  \begin{gathered}
		\includegraphics[height=1cm]{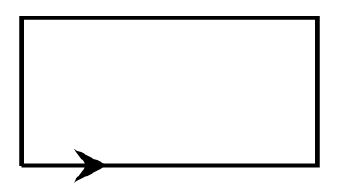} \vspace{-0.3cm}
	\end{gathered} \; \rangle = \langle \; \begin{gathered}
		\includegraphics[height=1cm]{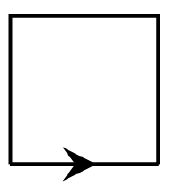} \vspace{-0.3cm}
	\end{gathered} \; \rangle^2
\end{align}

Similarly, all the Wilson loops in 2D lattice will factorize into Wilson loops within a single plaquette, which can be denoted as:
\begin{equation}
	w(n) = \langle \frac{1}{N}{\rm Tr}(U_P^{n}) \rangle \,,
\end{equation}
where $n$ represents the number that the loop winds around the plaquette, as shown below:
\begin{align}\label{factorize_1}
	w(1) = \langle \; \begin{gathered}
		\includegraphics[height=1cm]{figures/sec_2_fig/w1x1.pdf} \vspace{-0.3cm}
	\end{gathered} \; \rangle, \qquad 
	w(2) = \langle \; \begin{gathered}
		\includegraphics[height=1cm]{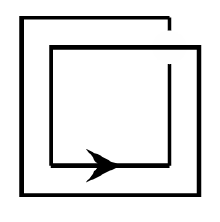} \vspace{-0.3cm}
	\end{gathered} \; \rangle, \qquad
	w(3) = \langle \; \begin{gathered}
		\includegraphics[height=1cm]{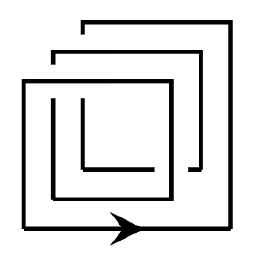} \vspace{-0.3cm}
	\end{gathered} \; \rangle, \quad \dots
\end{align}
In general, all $w(n)$ may appear after factorization, as we can see in the following example:
\begin{align}\label{factorize_2}
	\langle \;  \begin{gathered}
		\includegraphics[height=1cm]{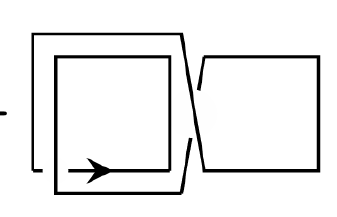} \vspace{-0.3cm}
	\end{gathered} \; \rangle = \langle \; \begin{gathered}
		\includegraphics[height=1cm]{figures/sec_2_fig/w1x1.pdf} \vspace{-0.3cm}
	\end{gathered} \; \rangle \times
	\langle \; \begin{gathered}
		\includegraphics[height=1cm]{figures/sec_2_fig/w_2.pdf} \vspace{-0.3cm}
	\end{gathered} \; \rangle
	= w(1) \times w(2)
\end{align}


If we further consider the Dyson-Schwinger equations \cite{Anderson_2017,Makeenko:1979pb}, we will find that all $w(n)$ with $n \geq 2$ can be solved as a polynomial of $w(1)$ and t'Hooft coupling constant $\lambda$ in the limit of $N \to \infty$. For example, $w(2) = 1-2\lambda\times w(1)$.

The solution of plaquette $w(1)$ is also known when $N \to \infty$ \cite{PhysRevD.21.446,Wadia:2012fr}:
\begin{equation}
	w(1)=\left\{
	\begin{aligned}
		& 1-\frac{\lambda}{2} ,\quad 0<\lambda \leq 1 \\
		& \; \; \; \; \frac{1}{2\lambda}, \;\;\; \quad \lambda >1 	\end{aligned}
	\right. , \label{eq:plaquette}
\end{equation}
In subsequent calculations, we will not need the specific form of $w(1)$, but rather express all Wilson loops as polynomials of $w(1)$ and $\lambda$, we will also abbreviate $w(1)$ as $u$ in the rest of the paper.

\section{Conventions and Setup}\label{sec:conventions_and_setup}
In this section, we will prepare the dataset for the training and
introduce how we tokenize the Wilson loop's shape information and expectation value so that it can be taken as input of the Transformer.

\newcommand{\ew}{\langle \mathcal{W} \rangle}

\subsection{Prepare dataset}

We take all Wilson loops with length less than or equal to 16 as the dataset. Since Wilson loops are invariant under the lattice symmetry group \cite{Kazakov:2024ool,Guo:2025fii,Li:2024wrd}, conjugate transformation and cyclic permutation, we collect all equivalent samples of a certain Wilson loop as an ``equivalence class''. We show the number of samples and equivalence classes in Table \ref{tab:sample_number}.
\begin{table}[H]
	\centering
	\renewcommand{\arraystretch}{1}
	\begin{tabular}{
			cccccccc
		}
		\toprule[1.5pt]
		Length & 4 & 6 & 8 & 10 & 12 & 14 & 16  \\
		\midrule[1pt]
		Samples & 8 & 24 & 216 & 1520 &  12080 & 94808 & 762104 \\
		Class  & 1  & 1 & 7 & 15 &  95 & 456 &  3217 \\
		\bottomrule[1.5pt]
	\end{tabular}
	\caption{
		\label{tab:sample_number}
		The number of samples for different lengths.
	}
\end{table}

After obtaining all the shape information of Wilson loops, we use the factorization property of 2D lattice YM theory and apply the method introduced in last section to express their expectation as the polynomial of $\lambda$ and $u$.

Before proceeding with the dataset preparation, it is important to characterize the structure of the output space. We have systematically checked all analytic expressions for Wilson loops with length $L \leq 16$ and found that they are constructed from only $10$ basic factors:
\begin{align}
	&
	u,\,
	u^2-2,\,
	u-2 \lambda ,\,
	2 \lambda  u-1,\,
	4 \lambda  u-1,\,
	u^2+2 \lambda  u-2,\,
	\\ \nn
	&
	u^2+4 \lambda  u-2,\,
	2 u^2+6 \lambda  u-3,\,
	4 \lambda ^2-6 \lambda  u+1
	\,.
\end{align}
These $10$ factors combine to give only $46$ distinct polynomial forms:
\begin{align}
	& u^{\alpha}~~(\alpha=1, \ldots ,16), \nn\\
	& u^{\alpha} (1-2 \lambda  u)~~(\alpha=0, \ldots ,8), ~~~~~~~~~~~~u^\alpha( 1-2 \lambda  u)^2~~(\alpha=0,1,2), \nn \\
	& u^{\alpha}\left(2-u^2\right)~~(\alpha=2,3,4) , ~~~~~~~~~~~~~~~u^\alpha(u-2 \lambda ) (1-2 \lambda  u)~~(\alpha=0,1,2,3),\nn\\
	& u^\alpha \left(2 - 2\lambda u - u^2\right)~~(\alpha=2,3,4), ~~~~~~u^\alpha \left(2 - 4\lambda u - u^2\right)~~(\alpha=2,\dots,6),\nn\\
	& u^5 \left(3 - 6\lambda u - 2u^2\right), ~~\quad u (1-2 \lambda  u) (1-4 \lambda  u), ~~\quad
	(1-2 \lambda  u) \left(1+4 \lambda ^2-6 \lambda  u\right).
\end{align}
This restricted output space—only $46$ distinct polynomial forms built from $10$ basic factors—significantly simplifies the learning task and is a key characteristic of the 2D exactly-solvable theory.

After all the dataset are generated, we are ready to conduct learning experiments.  For all the experiments, we will choose the training set, test set, and validation set from each equivalence class with the same proportion.

\subsection{Conventions}
Above we generate all the Wilson loops for the training and classify them into different equivalent class. In this subsection we introduce how to translate the shape and expectation value of a Wilson loop into a sequence which can be taken as inputs of Transformer.

\begin{figure}[t]
	\begin{center}
		\includegraphics[width=0.5\textwidth]{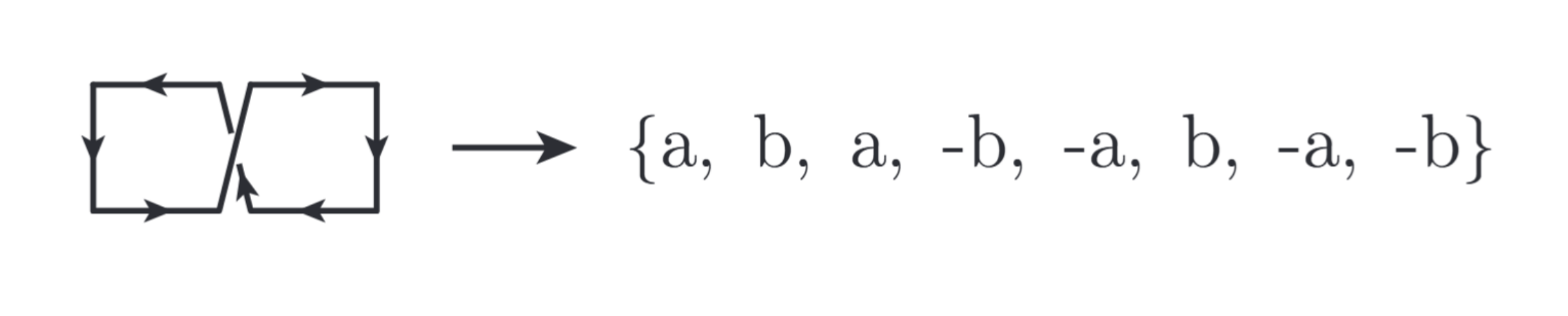}
	\end{center}
	\caption{A Wilson loop of length 8.}\label{fig:wilson_loop_length_8_example}
\end{figure}

Recall that Wilson loops are gauge invariant operators constructed by traces of products of links $U_\mu$. In 2 dimensions, we establish a mapping between links and letters:
\begin{equation}
	U_{\mu_0} \rightarrow \text{a},\ U_{\mu_1} \rightarrow \text{b} \ ,
\end{equation}
with their Hermitian conjugates corresponding to
\begin{equation}
	U_{\mu_0}^\dagger \rightarrow \text{-a},\ U_{\mu_1}^\dagger \rightarrow \text{-b} \ .
\end{equation}
This symbolic representation compactly encodes the shape information of Wilson loops. The plaquette, for instance, can be translated into a sequence of letters as $\{\text{a},\ \text{b},\ \text{-a},\ \text{-b}\}$. Another example is given in the  Figure~\ref{fig:wilson_loop_length_8_example}, where a more complex Wilson loop of length 8 is given. The length of a Wilson loop is defined as the number of links it contains.
Obviously, by this notation, the shape information of Wilson loop is expressed as an ordered list of letters. Thus it has been translated to the language of Transformer. The input sequences have four tokens in total, they are `a', `-a', `b' and `-b' in this study.


Now we consider the output, i.e., the expectation values. Taking the Wilson loop $\{\text{a},\ \text{a},\ \text{b},\ \text{-a},\ \text{-b},\ \text{a},\ \text{b},\ \text{-a},\ \text{-a},\ \text{-b}\}$ as an example, its analytic expectation value takes the polynomial form $u - 2 \lambda u^2$, where $u$ denotes the expectation value of plaquette as shown in eq.\eqref{eq:plaquette}. In general, the expectation value is a polynomial of following form
\begin{equation}
	\ew = \sum_{i=1}^n S_i \times A_i \times \lambda^{B_i} \times u^{C_i} \ .~~~\label{w-form}
\end{equation}
$S_i \in \{-1,\ 1\}$ specifies the sign of the term. $A_i$ gives the absolute value of the coefficient. $B_i$ and $C_i$ count powers of $\lambda$ and $u$ respectively. Then, each term in eq.\eqref{w-form}  can be uniquely determined by given a quadruple $\{S_i,\ A_i,\ B_i,\ C_i\}$.  For example, with this mapping, the polynomial $u - 2 \lambda u^2$ could be expressed as two quadruples
\begin{equation}
	u - 2 \lambda u^2 \rightarrow \{1,\ 1,\ 0,\ 1\},\ \{-1,\ 2,\ 1,\ 2\} \ .~~~\label{exa-1}
\end{equation}
Due to the monomials in a polynomial can be exchanged, the representation of the polynomial is still not unique. To ensure the uniqueness, we further impose a global ordering: first by $B_i$ in descending order, then by descending $C_i$. Finally we flatten the sorted tuples into a string format. The example eq.\eqref{exa-1} now gives
\begin{equation}
	\begin{aligned}
		\{1,\ 1,\ 0,\ 1\},\ \{-1,\ 2,\ 1,\ 2\} &\xrightarrow{\text{Sort}} \{-1,\ 2,\ 1,\ 2\},\ \{ 1,\ 1,\ 0,\ 1\} \\
		&\xrightarrow{\text{Flatten}} \text{`-',\ `2',\ `1',\ `2',\ `+',\ `1',\ `0',\ `1'}\ ,
	\end{aligned}
	\label{eq:flatten_example}
\end{equation}
where signs are encoded as ``-" or ``+". 

Now we can express both Wilson loops and their expectation values in a sequence form. We shall use this format to handle the Wilson loops in order to prepare datasets for the following experiments.

\section{Overview of learning result}
\label{sec:overview_of_learning_result}

With the preparation of the last section, we are able to use the Transformer to study the map between the shape information and expectation of a Wilson loop.
We have completed experiments with various dataset configurations and will present the learning curves in the following parts of this article. We adopt the standard encoder-decoder Transformer architecture as originally proposed in \cite{vaswani2023attentionneed}. The encoder processes the input Wilson loop sequence and the decoder autoregressively generates the output polynomial expression. The model dimension, number of attention heads, and number of layers (applied equally to both encoder and decoder) are treated as hyperparameters and will be discussed in section~\ref{sec:hyperopt}. The accuracy in the learning curve is obtained by testing the validation sets at the end of each epoch during training. One epoch is defined as training on 100,000 samples in the following experiments. The models are trained using the Adam optimizer with CrossEntropyLoss as the loss function. The learning rate is fixed at 0.001 without any learning rate scheduler. The batch size is set to 256 for all experiments. Here, accuracy is measured as the exact full-sequence match — that is, a prediction is considered correct only when the entire generated sequence matches the ground-truth expression exactly.






Before presenting the learning results, we analyze the label distribution and establish baseline performance. The $L \leq 16$ dataset contains 46 distinct polynomial forms with a total of 870,760 samples. Table~\ref{tab:l16_label_dist} shows the distribution of all forms sorted by frequency. The most common form, $u^4$, accounts for approximately $20.5\%$ of all samples. A naive baseline classifier that always predicts the most frequent form would achieve at most $\sim 20.5\%$ accuracy. 

\begin{table}[htbp]
	\centering
	\footnotesize
	\setlength{\tabcolsep}{3pt}
	\begin{tabular}{lrr}
		\toprule
		Polynomial Form & Sample Count & Percentage \\
		\midrule
		$u^4$ & 178,640 & 20.52\% \\
		$u^5$ & 169,560 & 19.47\% \\
		$u^6$ & 113,664 & 13.05\% \\
		$u^3$ & 96,352 & 11.07\% \\
		$u^7$ & 70,816 & 8.13\% \\
		$u^8$ & 43,384 & 4.98\% \\
		$u^3 - 2u^4\lambda$ & 34,176 & 3.92\% \\
		$u^9$ & 27,000 & 3.10\% \\
		$u^2 - 2u^3\lambda$ & 23,872 & 2.74\% \\
		$u^4 - 2u^5\lambda$ & 22,336 & 2.57\% \\
		$u^2$ & 18,968 & 2.18\% \\
		$u^{10}$ & 16,904 & 1.94\% \\
		$u^5 - 2u^6\lambda$ & 11,472 & 1.32\% \\
		$u^{11}$ & 10,144 & 1.16\% \\
		$u^6 - 2u^7\lambda$ & 5,632 & 0.65\% \\
		$u^{12}$ & 4,888 & 0.56\% \\
		$u - 2u^2\lambda$ & 4,336 & 0.50\% \\
		$2u^4 - u^6 - 4u^5\lambda$ & 3,024 & 0.35\% \\
		Other 28 forms & 5,952 & 0.68\% \\
		\midrule
		Total & 870,760 & 100\% \\
		\bottomrule
	\end{tabular}
	\caption{Distribution of polynomial forms in the $L \leq 16$ dataset. The top 18 forms account for $99.3\%$ of all samples. A naive baseline that always predicts the most common form ($u^4$) would achieve $\sim 20.5\%$ accuracy.}
	\label{tab:l16_label_dist}
\end{table}

\begin{figure}[t
	]
	\centering
	\setlength{\tabcolsep}{-3pt}
	\subcaptionbox{
		Sequence lengths of 12 or less.
		\label{fig:b12-per}
	}{
		\includegraphics[width=0.3\textwidth]
		{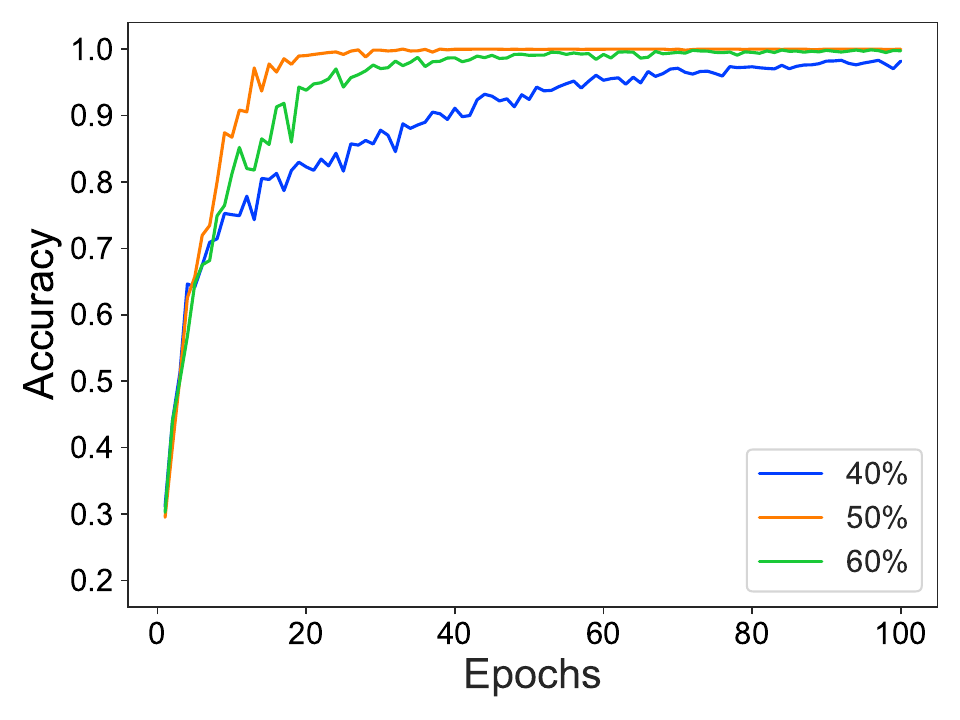}
	}
	\subcaptionbox{
		Sequence lengths of 14
		\label{fig:L14-per}
	}{
		\includegraphics[width=0.3\textwidth]
		{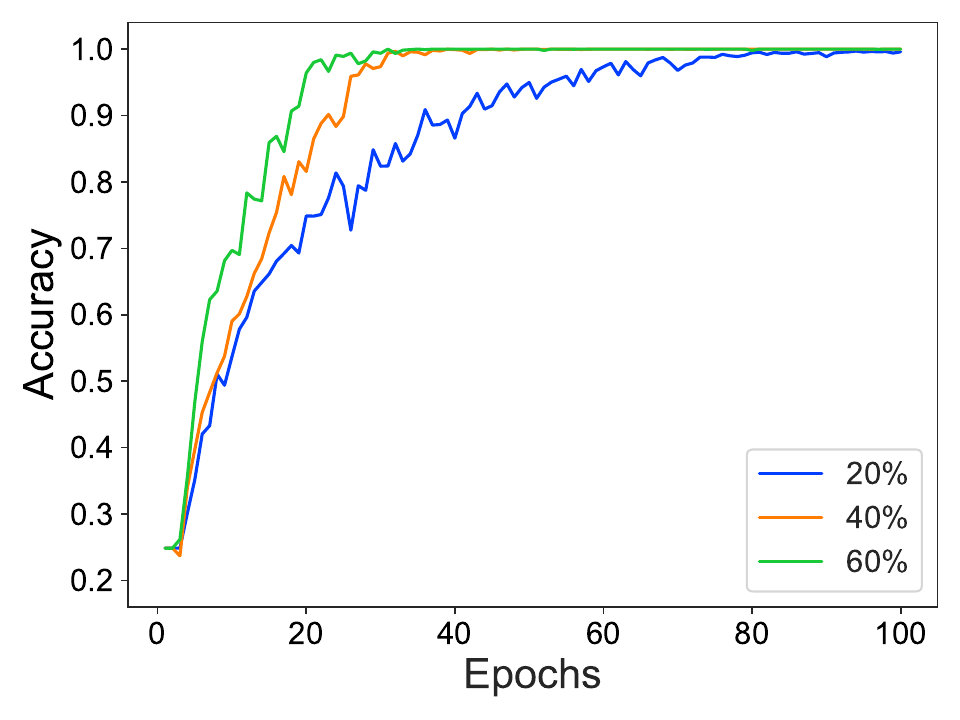}
	}
	\subcaptionbox{
		Sequence lengths of 16
		\label{fig:L16-per}
	}{
		\includegraphics[width=0.3\textwidth]
		{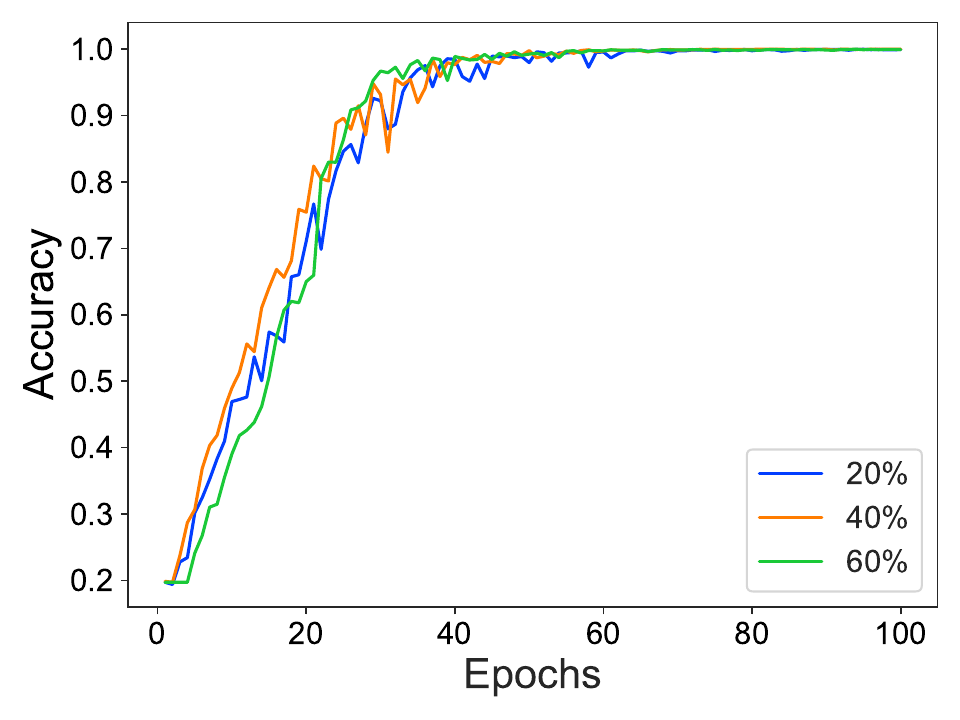}
	}
	\caption{The learning curves for different cases.}
	\label{fig:per-exp}
\end{figure}

With this label distribution in mind, we now present the learning results. In Figure~\ref{fig:per-exp}, we first present the best learning result of the model.
Specifically, we show the learning curves
for three cases: (a) sequence lengths of $12$ or less; (b) sequence lengths of $14$; (c) sequence lengths of $16$. 
For each length, we present three different curves. For case (a), the hyperparameters of Transformer are dimension $d = 128$, head $h = 8$, and layer $l = 4$. The three curves correspond to different proportions of training data over the whole data set,  which are $40\%$, $50\%$, and $60\%$, respectively. The model eventually achieved accuracy of $98.31\%$, $100.00\%$, and $99.88\%$ respectively. 
For case (b), the hyperparameters are dimension $128$, head $4$, and layer $4$. The three curves correspond to training data with proportion $20\%$, $40\%$, and $60\%$. They eventually achieved their highest accuracy of $99.70\%$, $100.00\%$, and $100.00\%$. For the case (c), the hyperparameters are dimension $128$, head $4$, and layer $4$. The three curves correspond to different proportions of training data over the whole data set,  which are $20\%$, $40\%$, and $60\%$, respectively.  They eventually achieved their accuracy of $99.97\%$, $99.99\%$, and $99.98\%$. 
Therefore, the Transformer's high accuracy   represents a substantial improvement over this baseline, demonstrating that the model learns to distinguish among the diverse polynomial forms rather than simply memorizing the most common one.

From these curves, one can see that our model has achieved high accuracy on the supervised mapping task, i.e., predicting the analytic expressions of vacuum expectation values of various Wilson loops from their shape information. 
We carefully choose the hyperparameters so that the model achieve best performance, we will discuss the impact of hyperparameters in next section.

	
\section{Impact of  hyperparameters }
\label{sec:hyperopt}

In this section, we focus on the impact of hyperparameters by using the Wilson loops with length $14$. 
The results will guide our choice of hyperparameters in our further training experiments later. Since we use Transformer as the model, the hyperparameter configurations explored in this study include:
\begin{itemize}
	\item {\bf Model dimensions} ($\dm$) with two choices:  $128$ and $256$;
	
	\item {\bf Attention heads} ($\nh$) with two choices: $8$ and $16$;
	
	\item {\bf Number of layers} ($\nl$) with three choices:  $4$, $6$, and $8$.
\end{itemize}
With this study, we are able to choose a proper parameter configuration by investigating the final accuracy and training duration.


Within the training, each model is trained over $100$ epochs. For a sequence with length $14$,
we have randomly chosen  $55,869$ samples as the training set, $1,307$ samples as the validation set, and the remaining $37,632$ samples as the test set. 

\begin{figure}[t]
	\centering
	\subcaptionbox{
		$\nh=8$, $\nl=4$
		\label{fig:L14-hyper-8-4}
	}{
		\includegraphics[width=0.3\textwidth]{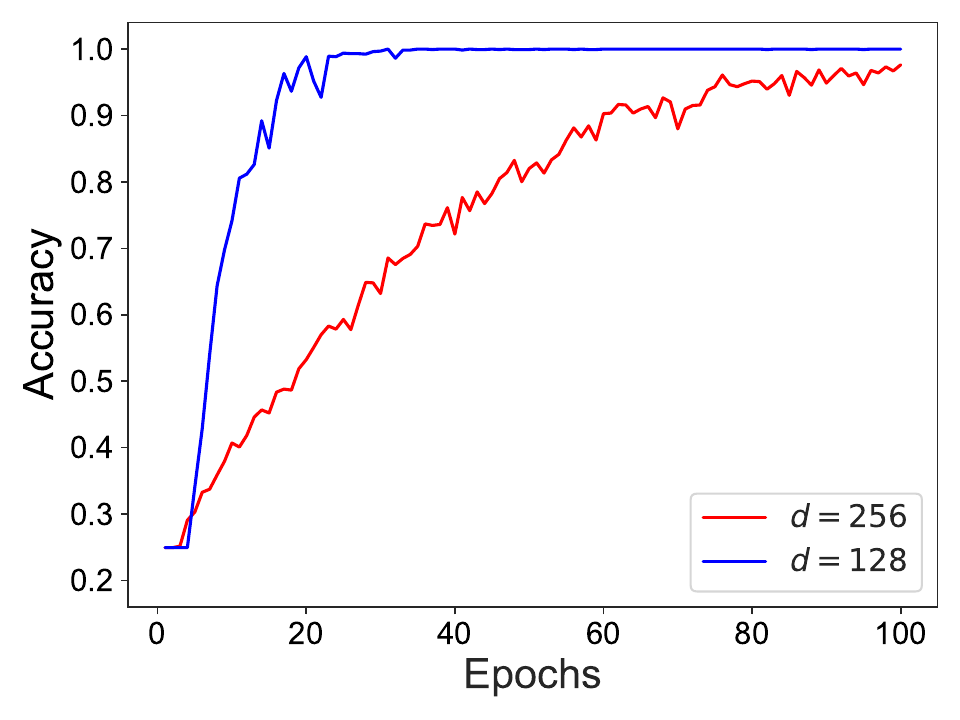}
	}
	\subcaptionbox{
		$\nh=8$, $\nl=6$ 
		\label{fig:L14-hyper-8-6}
	}{
		\includegraphics[width=0.3\textwidth]{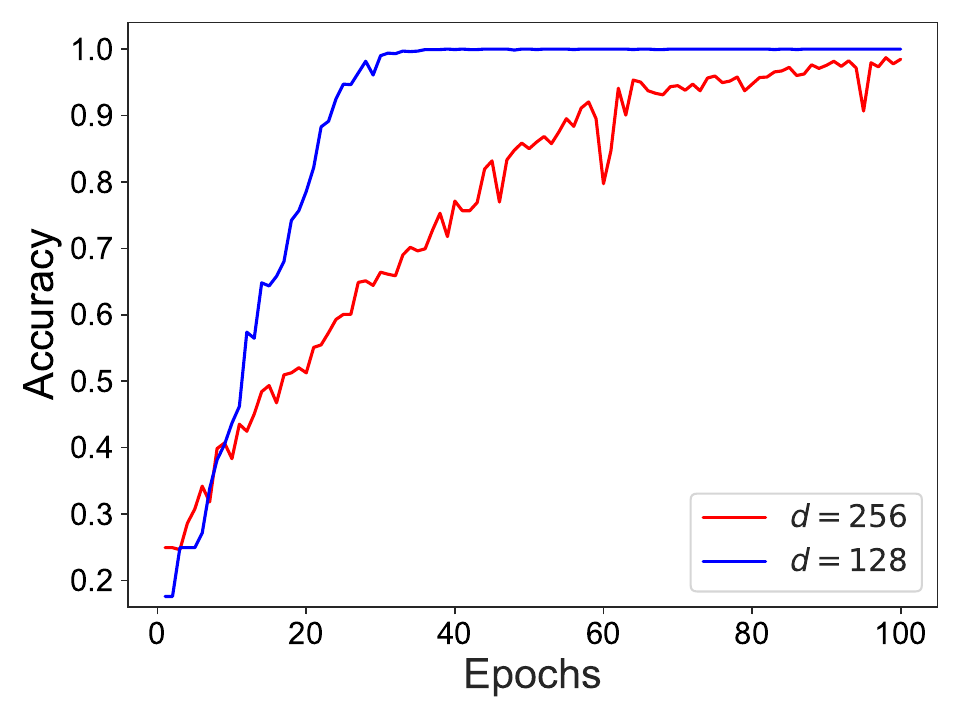}
	}
	\subcaptionbox{
		$\nh=8$, $\nl=8$ 
		\label{fig:L14-hyper-8-8}
	}{
		\includegraphics[width=0.3\textwidth]{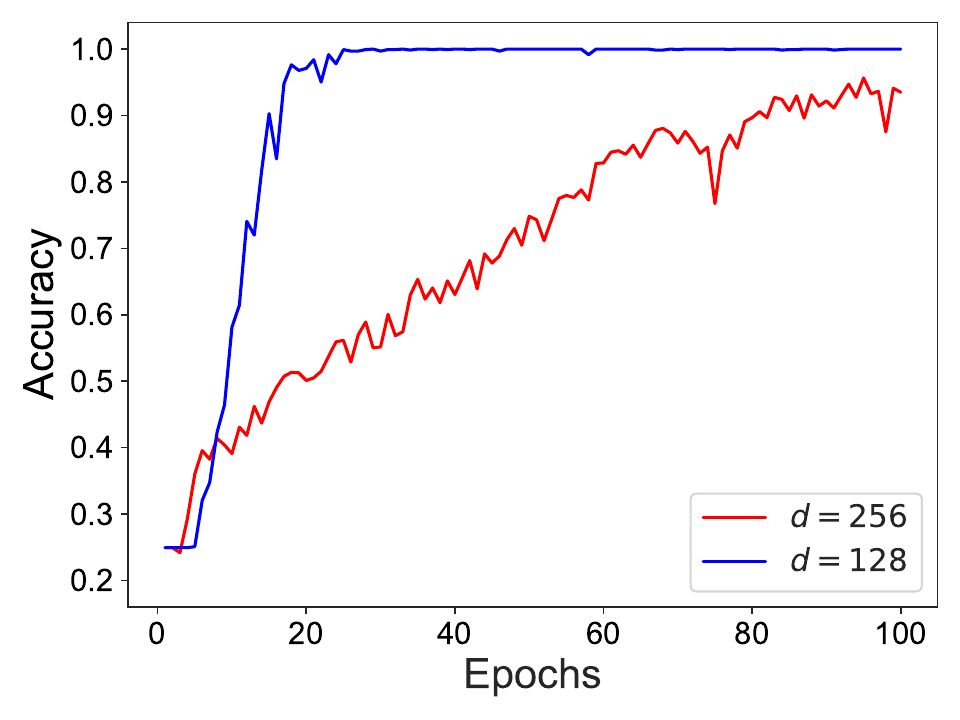}
	}
	\\
	\subcaptionbox{
		$\nh=16$, $\nl=4$
		\label{fig:L14-hyper-16-4}
	}{
		\includegraphics[width=0.3\textwidth]{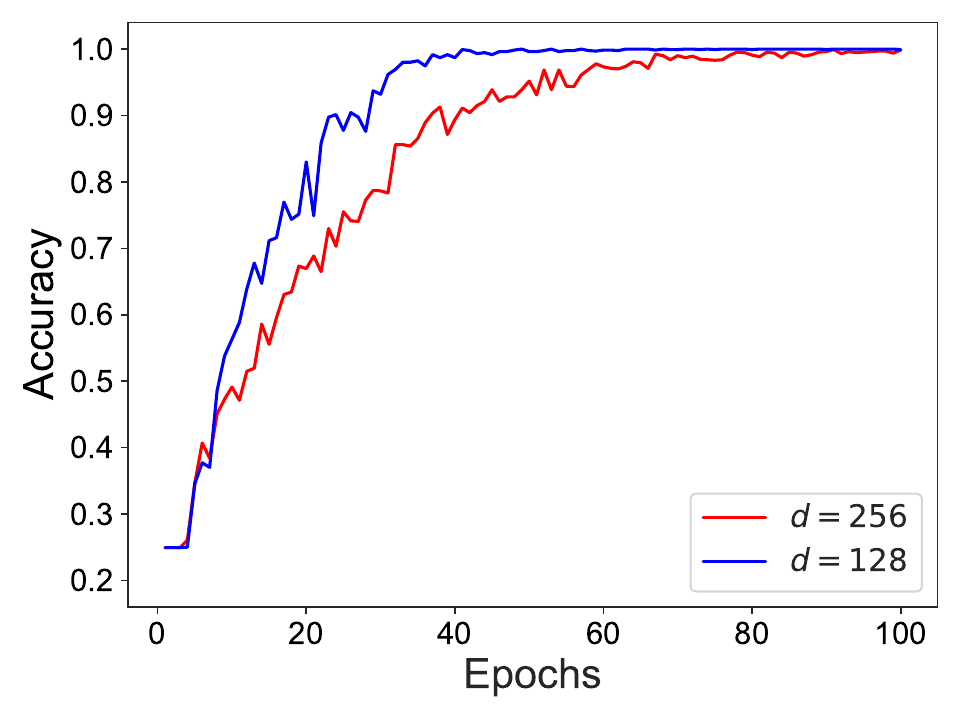}
	}
	\subcaptionbox{
		$\nh=16$, $\nl=6$ 
		\label{fig:L14-hyper-16-6}
	}{
		\includegraphics[width=0.3\textwidth]{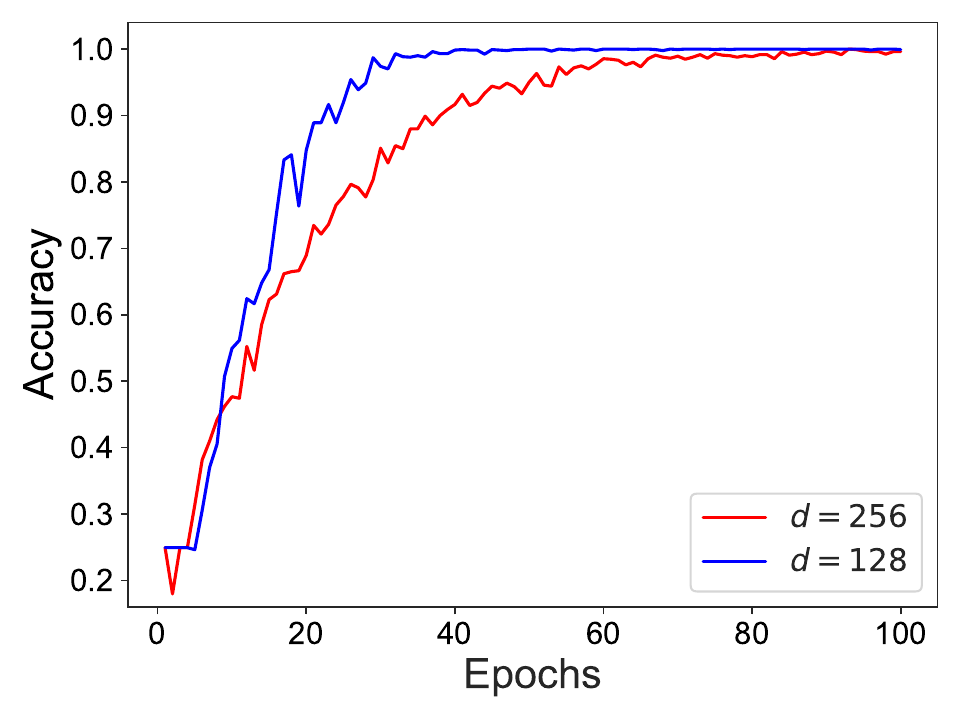}
	}
	\subcaptionbox{
		$\nh=16$, $\nl=8$ 
		\label{fig:L14-hyper-16-8}
	}{
		\includegraphics[width=0.3\textwidth]{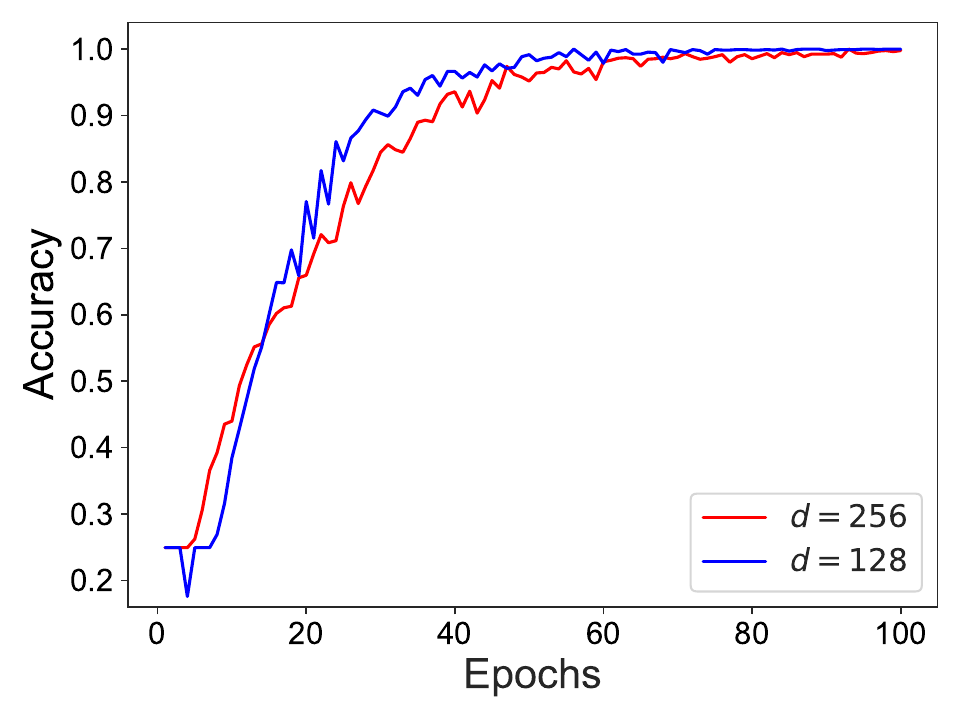}
	}
	\caption{The impact of the dimension. Each diagram has the same layer and head. In each diagram,   there are two learning curves corresponding to the dimensions $128$ and  $256$, respectively. }
	\label{fig:hyper-com-dim}
\end{figure}

We illustrate the impact of different hyperparameter configurations on several figures.

\begin{itemize}
	\item First, let us consider the influence of the dimension. In Figure \ref{fig:hyper-com-dim} we have presented the variation in accuracy across epochs with fixed $\nh$ and $\nl$, when we change the dimension from $128$ to $256$. Under the hyperparameters we chose, the model with dimension $128$ converges faster than the one with dimension $256$.
	
	\item Next, we consider the influence of the number of layers $\nl$. In Figure \ref{fig:hyper-com-layer}, we present the learning result for models with fixed $\dm$ , $\nh$ but different $\nl$. We observe that this parameter has relatively less impact on the model. The final accuracy and the convergence rate are almost the same except for the case with $\dm=256$ and $\nh=8$. For this one, the layer $8$ has the worst performance, both for the final accuracy and learning speed.
	
	\begin{figure}[t]
		\centering
		\subcaptionbox{
			$\dm=128$, $\nh=8$
			\label{fig:L14-hyper-128-8}
		}{
			\includegraphics[width=0.4\textwidth]{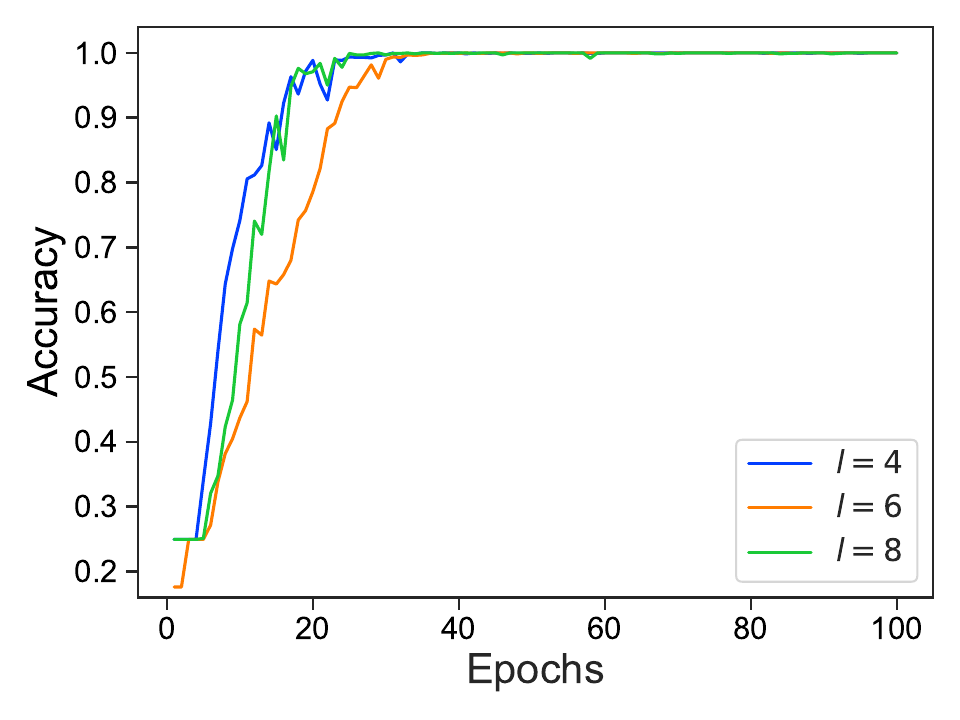}
		}
		\subcaptionbox{
			$\dm=128$, $\nh=16$
			\label{fig:L14-hyper-128-16}
		}{
			\includegraphics[width=0.4\textwidth]{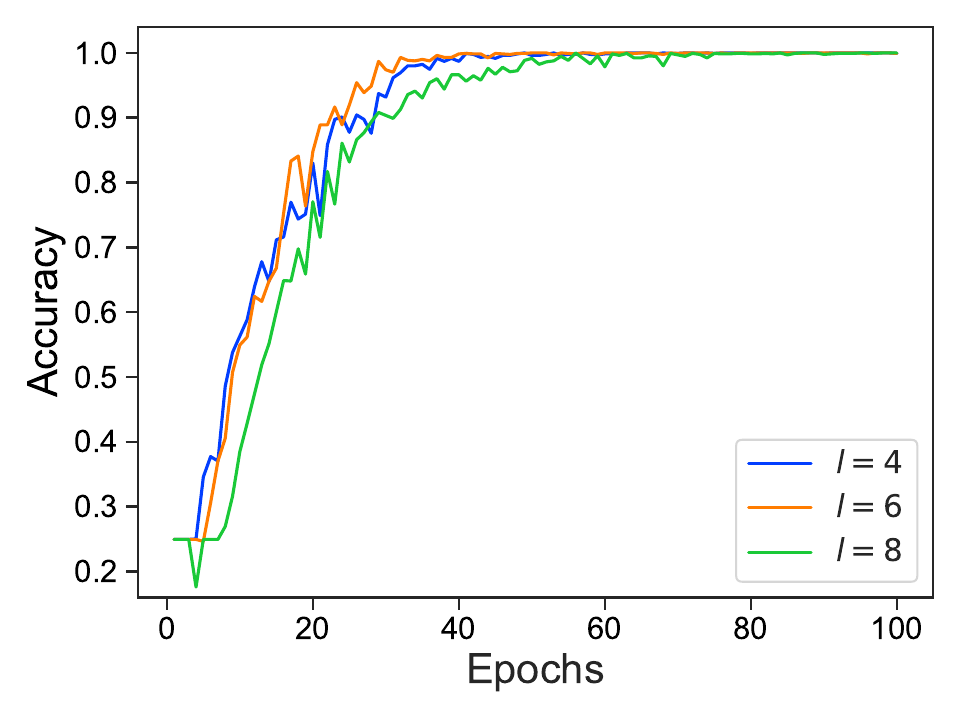}
		}
		\\
		\subcaptionbox{
			$\dm=256$, $\nh=8$
			\label{fig:L14-hyper-256-8}
		}{
			\includegraphics[width=0.4\textwidth]{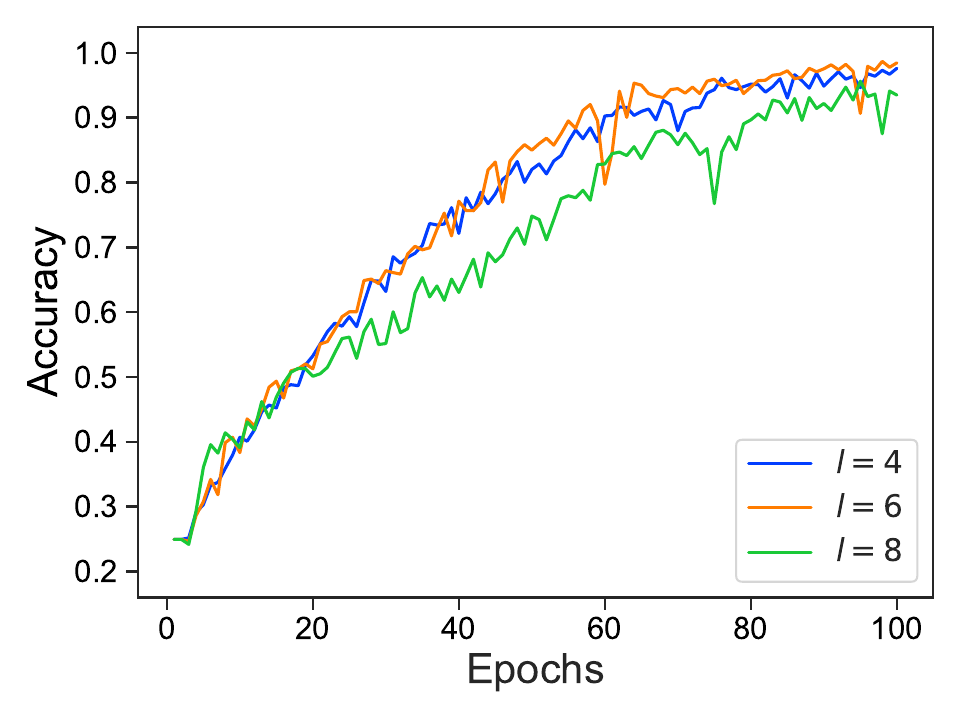}
		}
		\subcaptionbox{
			$\dm=256$, $\nh=16$
			\label{fig:L14-hyper-256-16}
		}{
			\includegraphics[width=0.4\textwidth]{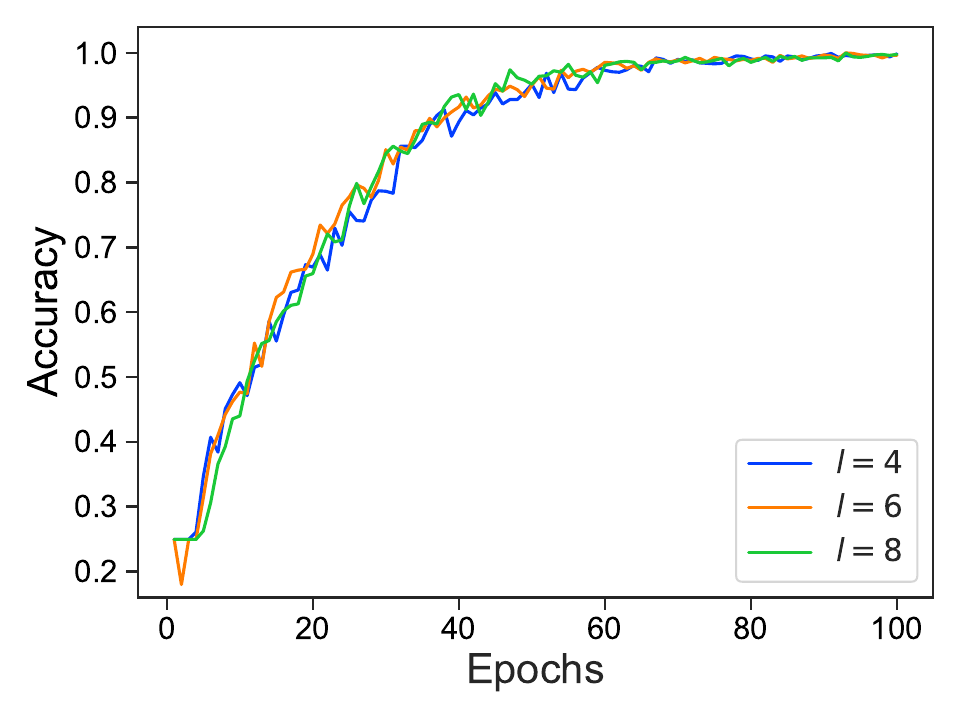}
		}
		\caption{The impact of $\nl$. Each diagram has the same dimension and head. In each diagram,   there are three learning curves corresponding to the layer $4$, $6$ and  $8$ respectively. }
		\label{fig:hyper-com-layer}
	\end{figure}
	
	\item Now, let us consider the influence of the head $\nh$. In Figure \ref{fig:hyper-com-head}, we have presented the variation in accuracy across epochs with fixed $\dm$ and $\nl$. From them, one can see that when $\dm = 128$, increasing the number of attention heads does not enhance the final accuracy. In fact, for layers $\nl = 4 $ or $8$, although models achieve the same performance, the convergence rate with head $16$ is worse than the one with head $8$.  
	
	\begin{figure}[h]
		\centering
		\subcaptionbox{
			$\dm=128$, $\nl=4$
			\label{fig:L14-hyper-128-4}
		}{
			\includegraphics[width=0.3\textwidth]{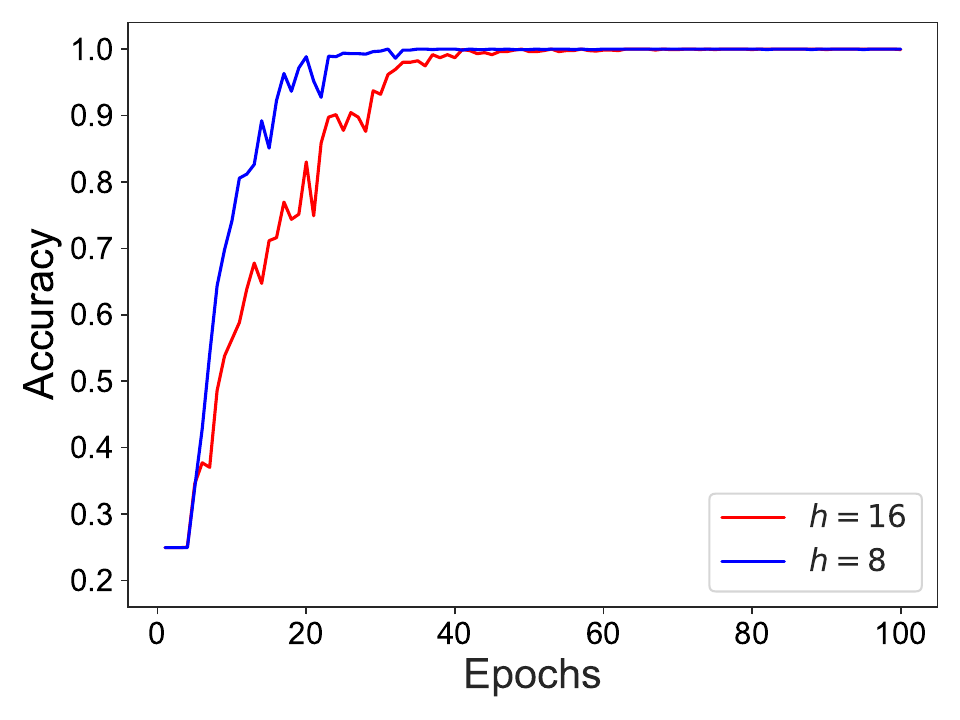}
		}
		\subcaptionbox{
			$\dm=128$, $\nl=6$
			\label{fig:L14-hyper-128-6}
		}{
			\includegraphics[width=0.3\textwidth]{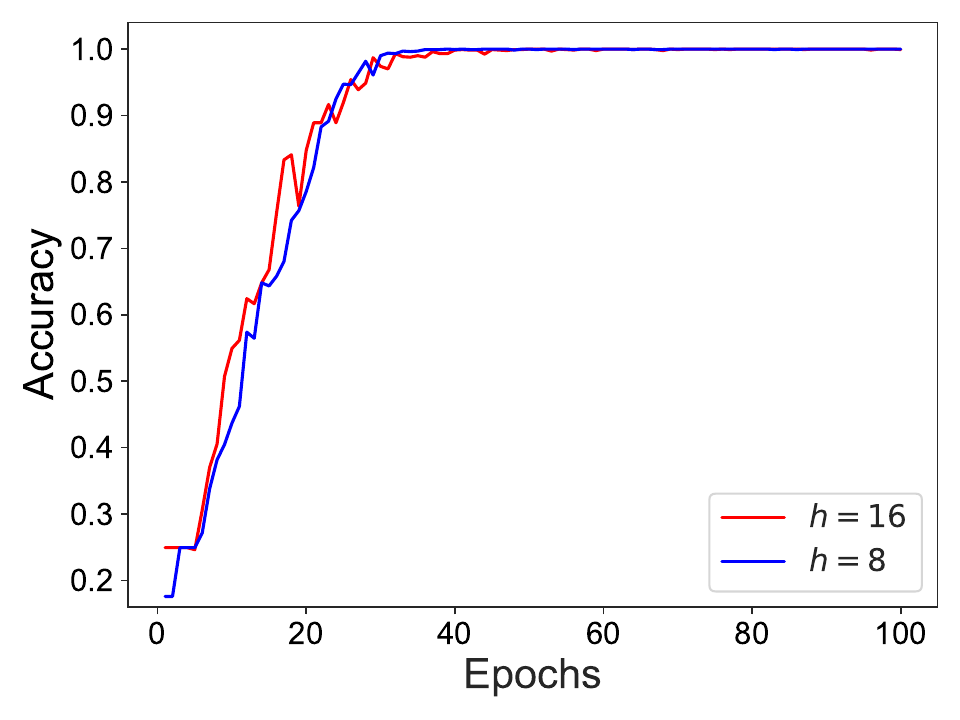}
		}
		\subcaptionbox{
			$\dm=128$, $\nl=8$
			\label{fig:L14-hyper-128-8h}
		}{
			\includegraphics[width=0.3\textwidth]{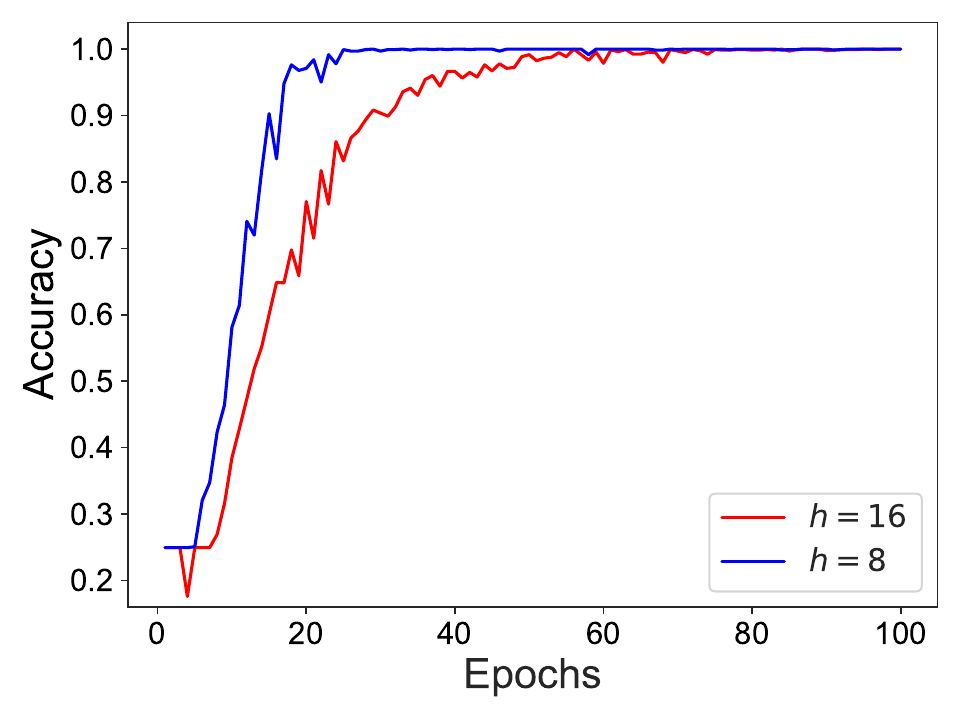}
		}
		\\
		\subcaptionbox{
			$\dm=256$, $\nl=4$
			\label{fig:L14-hyper-256-4}
		}{
			\includegraphics[width=0.3\textwidth]{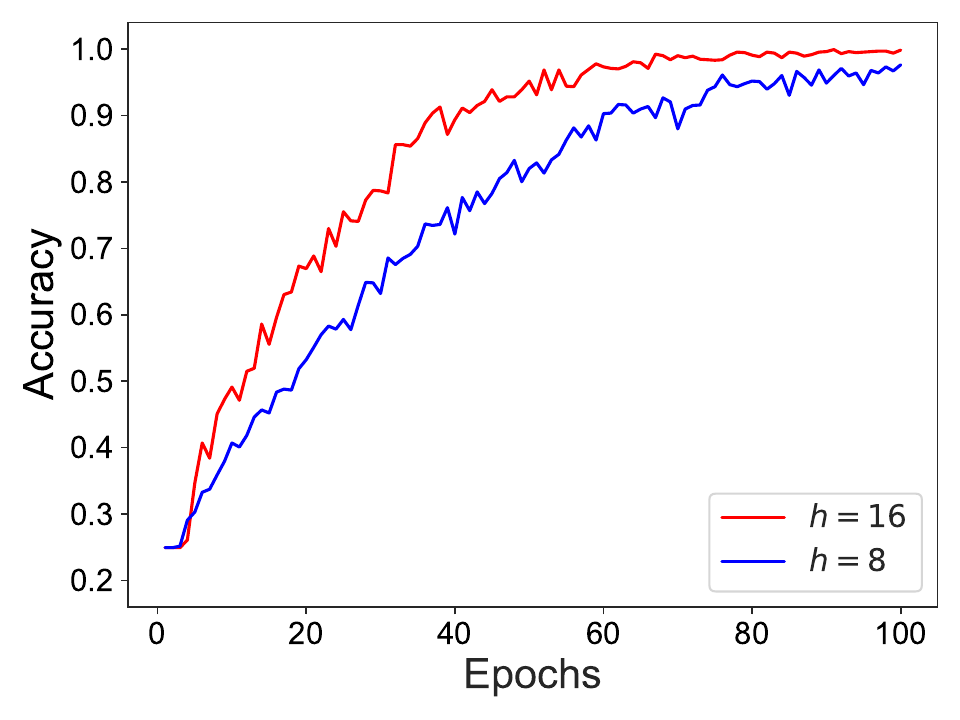}
		}
		\subcaptionbox{
			$\dm=256$, $\nl=6$
			\label{fig:L14-hyper-256-6}
		}{
			\includegraphics[width=0.3\textwidth]{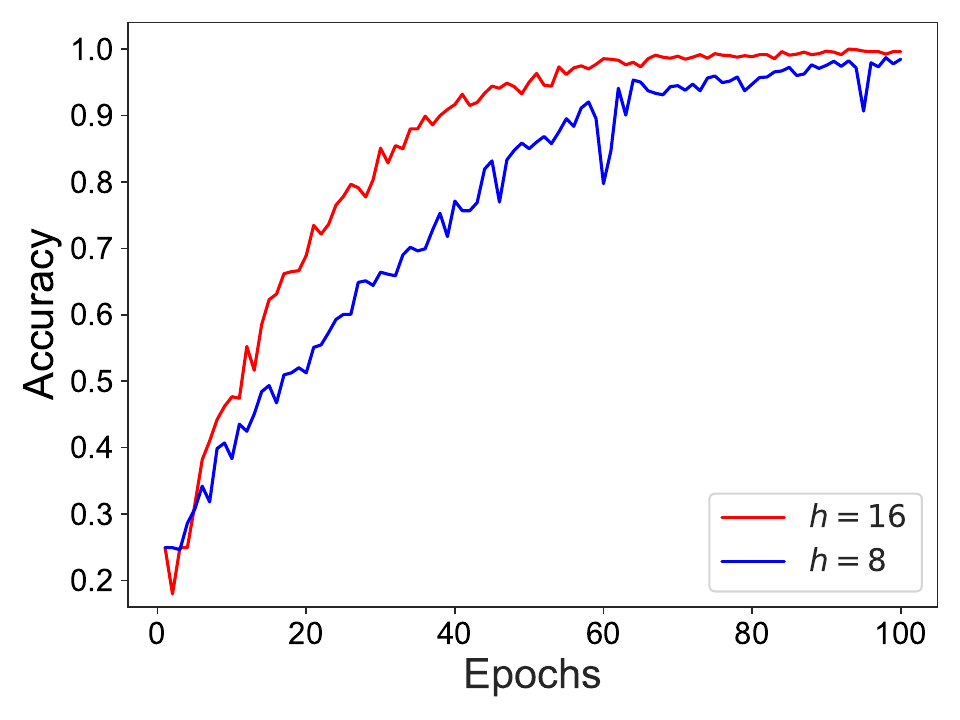}
		}
		\subcaptionbox{
			$\dm=256$, $\nl=8$
			\label{fig:L14-hyper-256-8h}
		}{
			\includegraphics[width=0.3\textwidth]{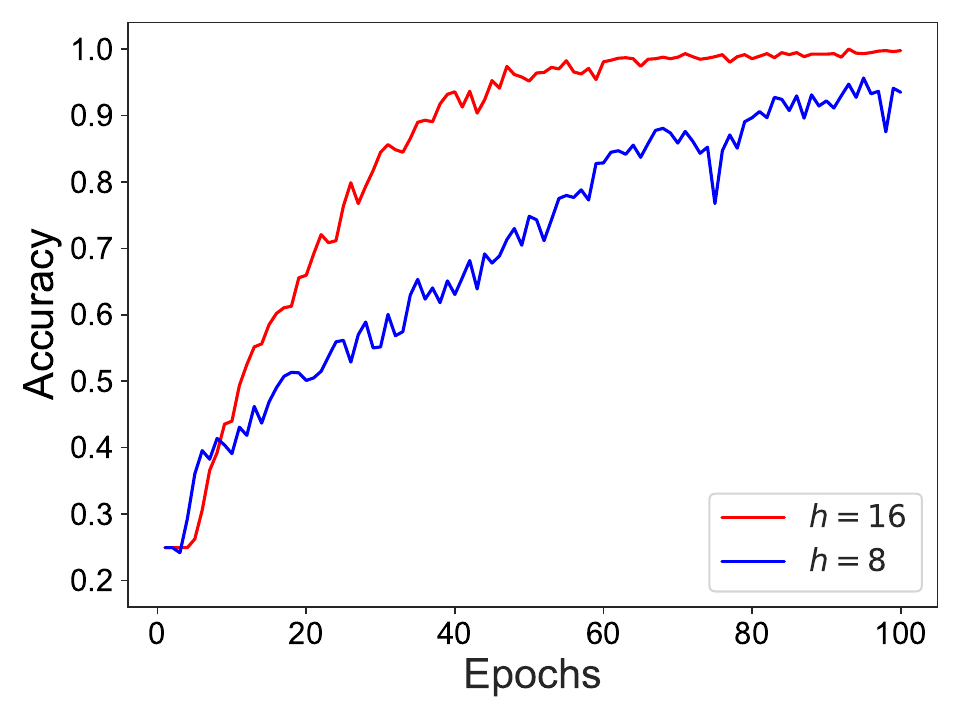}
		}
		\caption{The impact of $\nh$. There are two learning curves corresponding to the heads $8$ and $16$ respectively in each diagram.}
		\label{fig:hyper-com-head}
	\end{figure}
	
	In contrast, with $\dm = 256$, a higher number of attention heads will increase both the convergence rate and the final accuracy. This phenomenon implies that the head $h$ should increase synchronously with the growth of $d$.
	
	To further support this conclusion, we conducted supplementary experiments by training models with fixed dimension $\dm = 512$ and $\nl = 4$, but varying the number of attention heads with $8$, $16$, and $32$. The learning curves are shown in Figure~\ref{fig:512}.  One	sees that with attention heads $8$ and $16$, the model shows bad learning performance. However, increasing the number of attention heads to $32$ significantly enhances the model's learning performance. 
	
	\begin{figure}[t]
		\centering
		\includegraphics[scale=0.4]{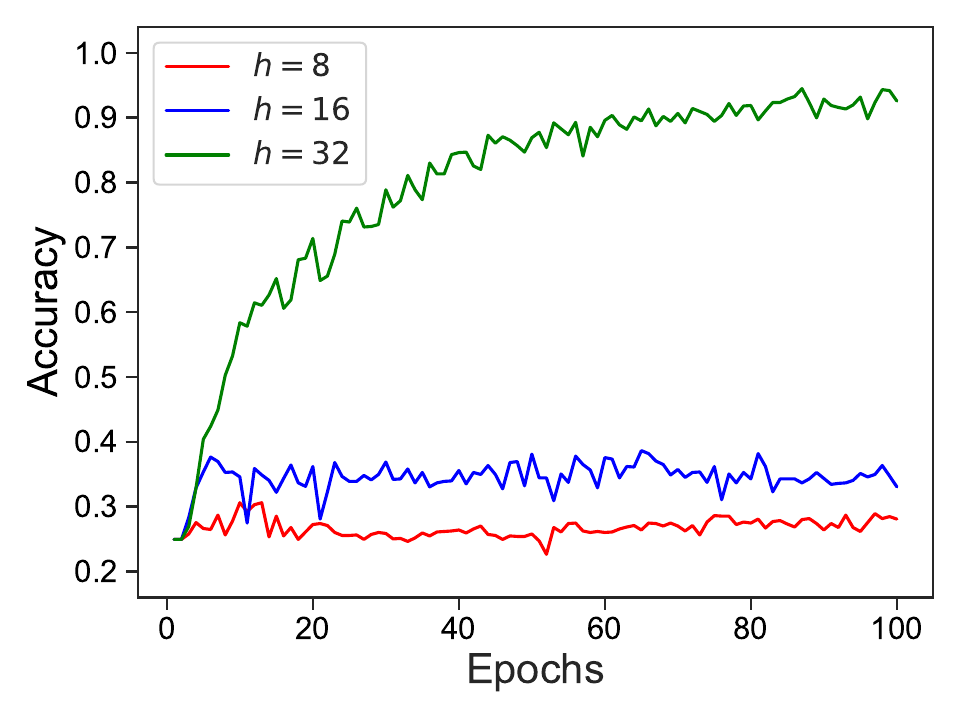}
		\caption{Further study of the impact of $\nh$. The model has the  fixed dimension $\dm = 512$ and number of layers $\nl = 4$.}
		\label{fig:512}
	\end{figure}
	
\end{itemize}

These hyperparameters should be appropriately balanced to achieve optimal performance.
Based on our experimental results,
we see that the optimal model configuration is with dimension $\dm = 128$,
layer $\nl = 4$,
and attention heads $\nh = 8$, which represents an effective balance between model capacity, convergence rate
and robustness. We will prefer to use these hyperparameters 
for subsequent experiments.
	
\section{Impact of training set proportion}
\label{sec:fractrain}
\begin{table}[t]
	\centering
	\renewcommand{\arraystretch}{1.5}
	\begin{tabular}{
			ccccccc
		}
		\toprule[1.5pt]
		Percentage & 30\% & 32.5\% & 35\% & 37.5\% & 40\% & 45\% \\
		\midrule[1pt]
		Train & 4087 & 4464 & 4800 & 5193 & 5482 & 6195 \\
		Test & 5674 & 5297 & 4961 & 4568 & 4279 & 3566 \\
		\bottomrule[1.5pt]
	\end{tabular}
	\caption{
		The specific number of data samples used in different percentages. The sizes of valid sets are all 4087.
	}
	\label{tab:sample_number-b12}
\end{table}

In model training, the proportion of training data over the whole data set is also a factor for the model's performance. In this section, we will examine how this proportion affects the model's learning efficiency.
We aim to identify the threshold of the proportion of the training set for the model to effectively learn the task and study the learning behaviours around the threshold.

For this training, we employ a dataset comprised exclusively of Wilson loops with lengths equal to or shorter than $12$. We have randomly 
generated six training sets with different proportions of the whole dataset as listed in Table \ref{tab:sample_number-b12}.

\begin{figure}[t]
	\centering
	\setlength{\tabcolsep}{-3pt}
	\subcaptionbox{30\% of the dataset\label{fig:b12-fraction-30}}{
		\includegraphics[width=0.4\textwidth]{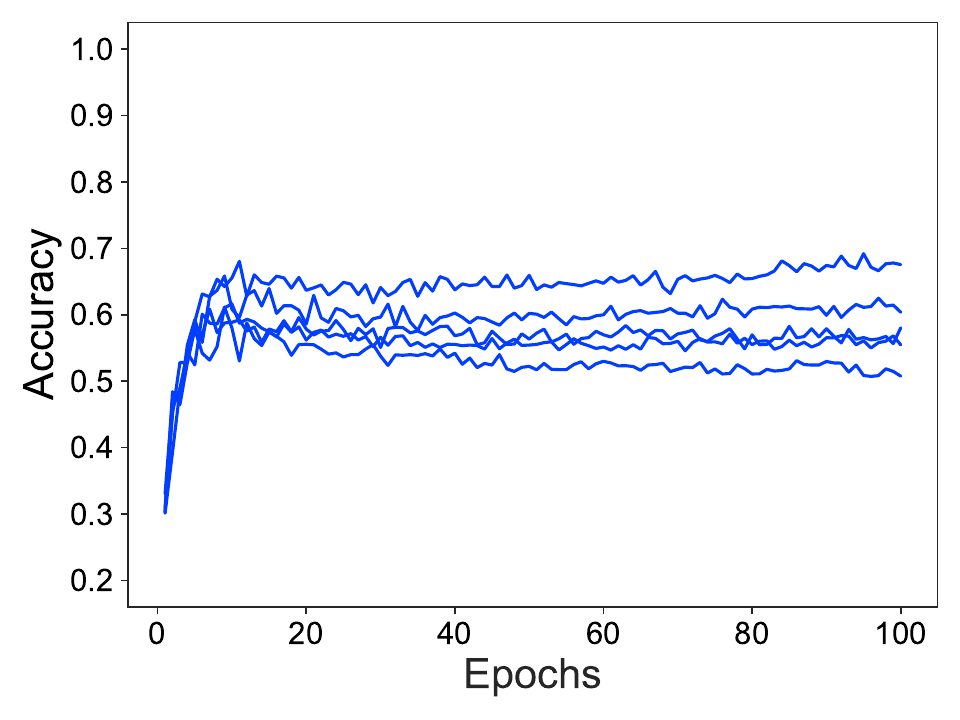}
	}
	\subcaptionbox{32.5\% of the dataset\label{fig:b12-fraction-32p5}}{
		\includegraphics[width=0.4\textwidth]{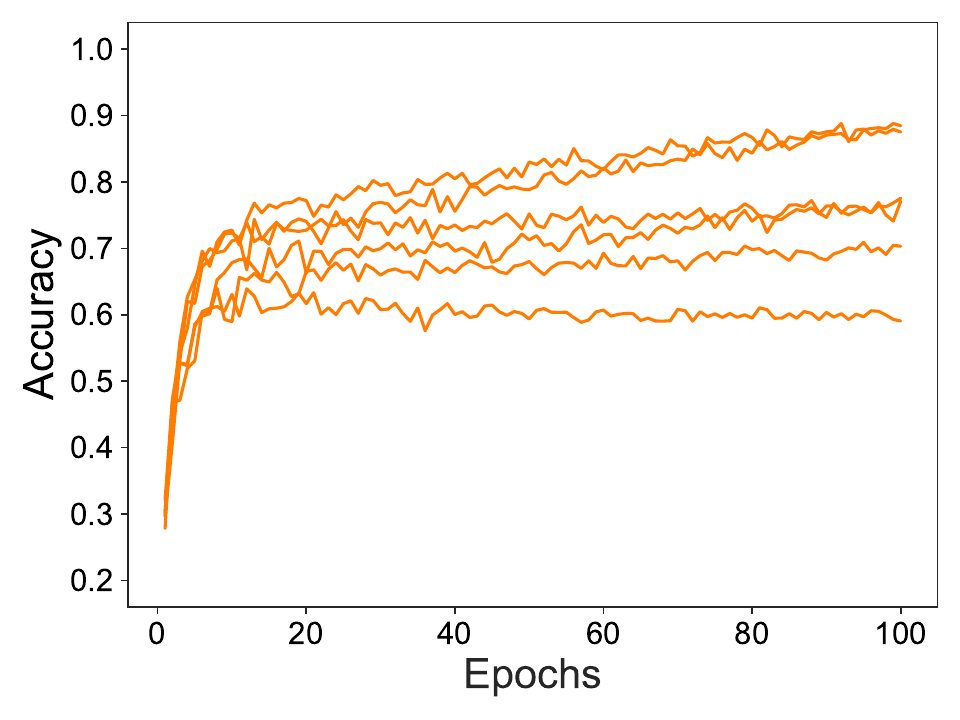}
	}
	\subcaptionbox{35\% of the dataset\label{fig:b12-fraction-35}}{
		\includegraphics[width=0.4\textwidth]{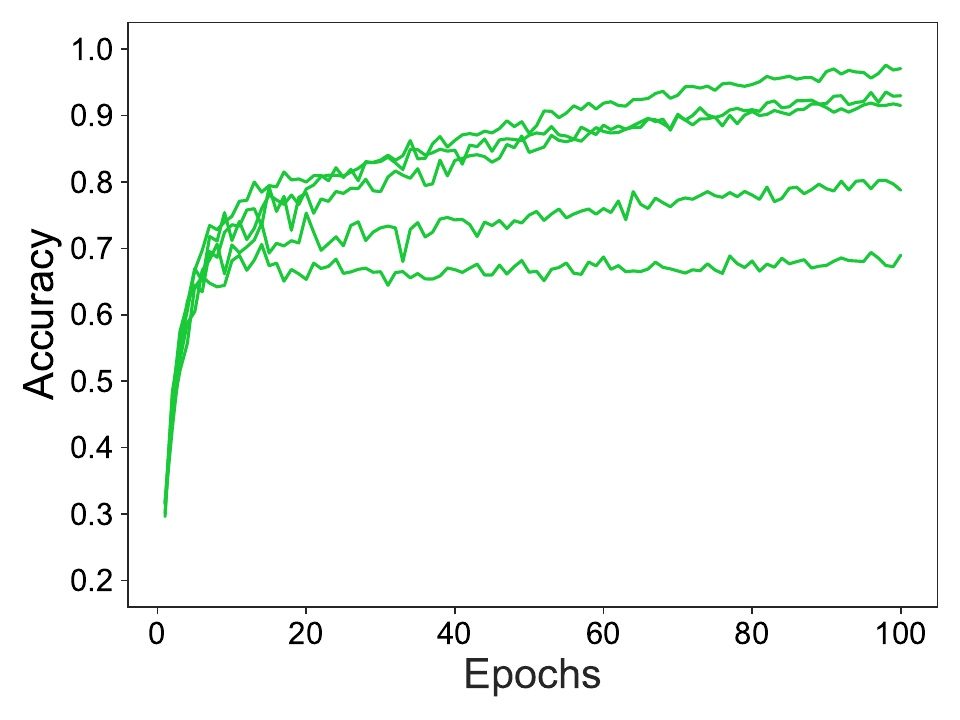}
	}
	\subcaptionbox{37.5\% of the dataset\label{fig:b12-fraction-37p5}}{
		\includegraphics[width=0.4\textwidth]{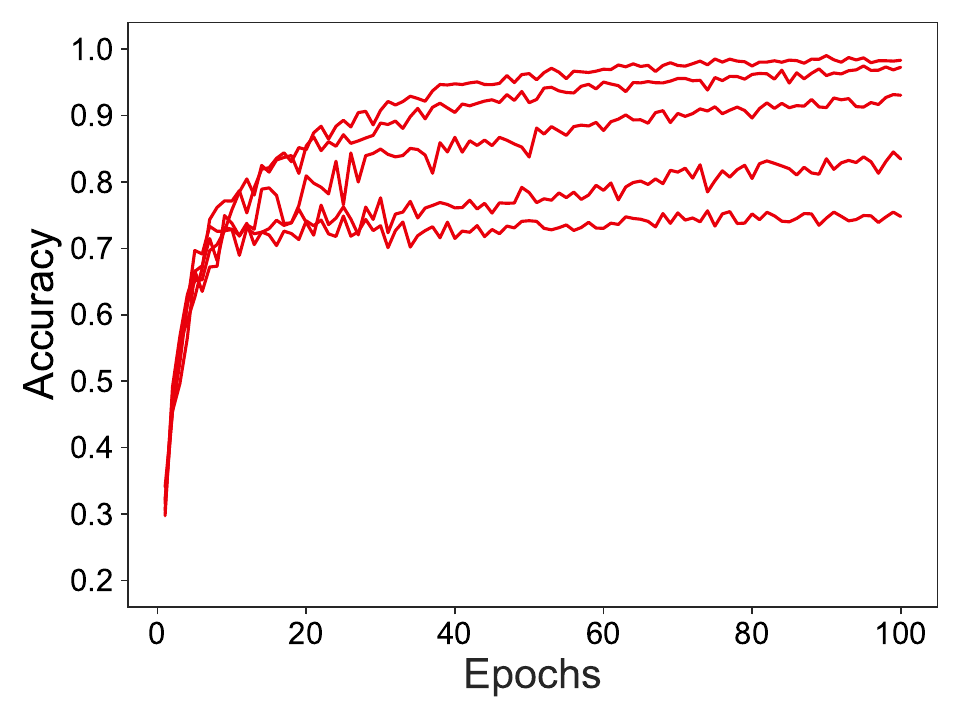}
	}
	\subcaptionbox{40\% of the dataset\label{fig:b12-fraction-40}}{
		\includegraphics[width=0.4\textwidth]{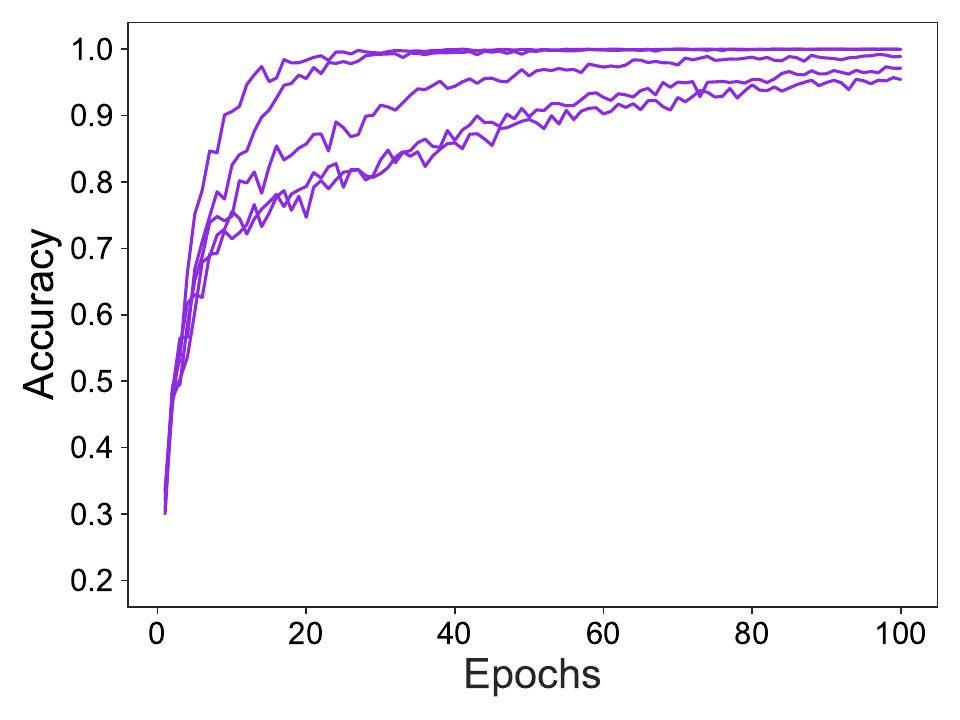}
	}
	\subcaptionbox{45\% of the dataset\label{fig:b12-fraction-45}}{
		\includegraphics[width=0.4\textwidth]{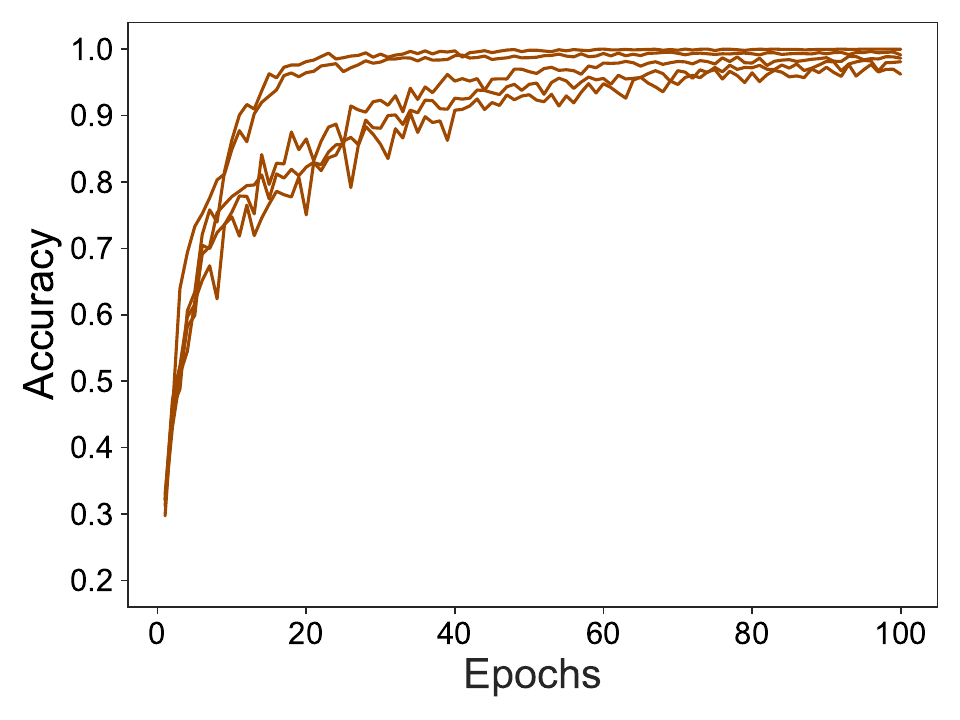}
	}
	\caption{Learning curves for various proportions of training dataset and  different initial condition of models. 
		For models in each sub picture, we randomly generate 5 different initial parameters and repeat the training.}
	\label{fig:b12-fraction-100}
\end{figure}

\begin{figure}[t]
	\centering
	\includegraphics[scale=0.5]{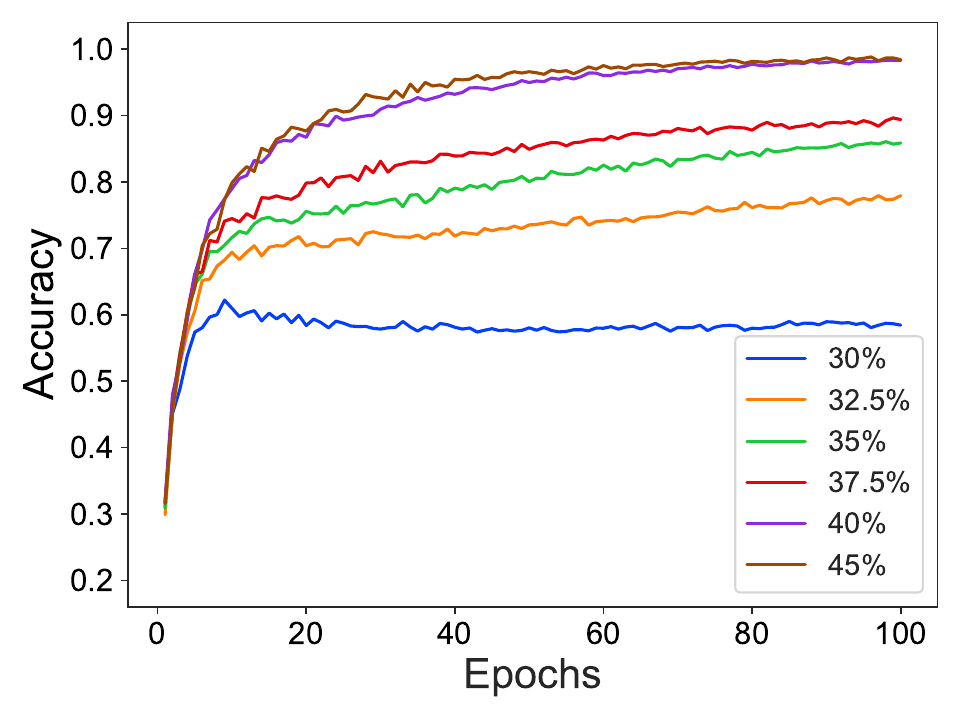}
	\caption{
		Learning curves for various proportions of training dataset. 
		Each line represents the mean value of five experiments with different initial condition.
	}
	\label{fig:L16-fraction-100-mean}
\end{figure}

The training results summarized in the 
Figure \ref{fig:b12-fraction-100} shows the model's accuracy across the initial 100 epochs. 
As it is well-recognized that initial conditions of model can significantly affect the learning efficiency, we conducted five experiments for each training set using different initial parameters\footnote{Distinguished from hyperparameters, these initial parameters refer to the model's learnable parameters.} to show potential biases. 
Specifically, we randomly generated five sets of initial parameters by uniform distribution and repeat the learning processes with these initial parameters. 

The results demonstrate that larger datasets contribute to convergence rate. 
For the experiment with proportions of $30\%$, no matter which initial parameters has been taken, it fails to learn the task to a high accuracy. 
For the experiments with proportions of $32.5\%$, $35\%$, and $37.5\%$, situations are different: some initial conditions have a good learning curve, while some initial conditions perform worse. 
For the model with proportions of 40\% and 45\%, although different initial conditions lead to different learning speeds, they all achieve a high learning accuracy. 
These three different behaviors lead us to claim that the model can not learn the task well with proportions of $30\%$ and below, while it succeeds with proportions of $40\%$ or above. 
The persistent upward trajectories observed in the $32.5\%$-$37.5\%$ curves prevent us from making definitive claims about their ultimate convergence to comparable accuracy levels.
Consequently, subsequent analysis will provide a more comprehensive examination of scenarios involving test set proportions between $30\%$-$37.5\%$. 
More tests and discussions about the impact of the initial conditions are given in \ref{sec:dataset-randomness}. In Fig \ref{fig:L16-fraction-100-mean}, we also show the average of accuracy for each sub figure of  Fig \ref{fig:b12-fraction-100}.

\begin{figure}[t]
	\centering
	\includegraphics[scale=0.45]{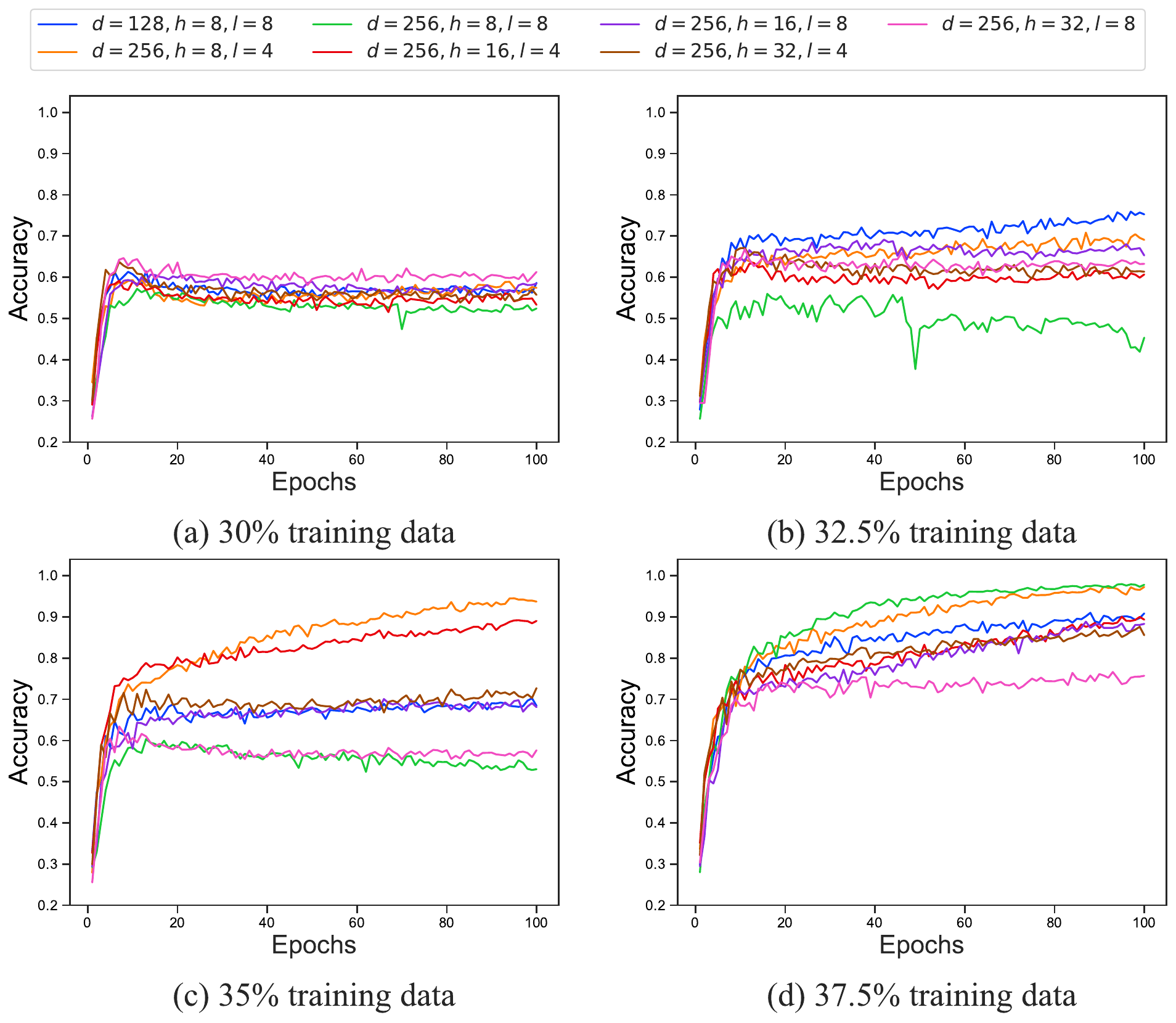}
	\caption{
		Learning curves of models with varying hyperparameters  trained on different dataset proportions (30\%, 32.5\%, 35\%, and 37.5\%).}
	\label{fig:b12-fraction-para}
\end{figure}

To further investigate the impact of the size of the training set on learning efficiency, we consider various combinations of 
hyperparameters of the model,  using the four dataset proportions that previously did not yield successful learning. 
The result is given in the Figure \ref{fig:b12-fraction-para}. Each diagram has multiple learning curves with hyperparameters indicated in the order of dimension, head, and layer. We can observe that the learning results in the Figure \ref{fig:b12-fraction-para} (c) exhibit the largest disparities among different hyperparameter selections. This indicates that near the threshold region, the choice of hyperparameters could also more significantly alter the learning outcomes. These conclusions are also used to guide our further tests.


\section{Enhancing Model Learning Ability through Mixed-Length Training}
\label{sec:mixtrain}

In this section, we will study the following problem: when we learn the model for evaluating the Wilson loops of a fixed length, would the prediction
be enhanced if adding information from shorter Wilson loops? If so, this would suggest that the Transformer can leverage shared structures between shorter and longer Wilson loops to improve its predictions. To check this idea, we add some data with length $14$ to
the training data with length $16$ and check the performance.

To make the mixed datasets meaningful, we examined the training dynamics of datasets with a sequence length of 16 under various training set proportions. In the Table \ref{tab:sample-number-l16} we list the number of data samples corresponding to various sampling percentages. For length 16, there are 762,104 samples in total. The training datasets
and the validation datasets  are constructed by randomly selected from each equivalence classes
in equal proportions.

\begin{table}[t]
	\centering
	\renewcommand{\arraystretch}{1.5}
	\begin{tabular}{
			ccccccc
		}
		\toprule[1.5pt]
		Percentage & 2.70\% & 3.12\% & 3.90\% & 4.27\% \\
		\midrule[1pt]
		Train & 20600 & 23815 & 29747 & 32511 \\
		Valid & 74210 & 74210 & 74210 & 74210 \\
		Test & 667294 & 664079 & 658147 & 655383 \\
		\bottomrule[1.5pt]
	\end{tabular}
	\caption{
		Table of the number of data samples in different percentages.
	}
	\label{tab:sample-number-l16}
\end{table}

To verify whether low-length datasets can enhance the learning of high-length data, we first determine the threshold for the high-length dataset and then examine whether the model can breakthrough this threshold after adding low-length data.

\begin{figure}[t]
	\centering
	\includegraphics[scale=0.5]{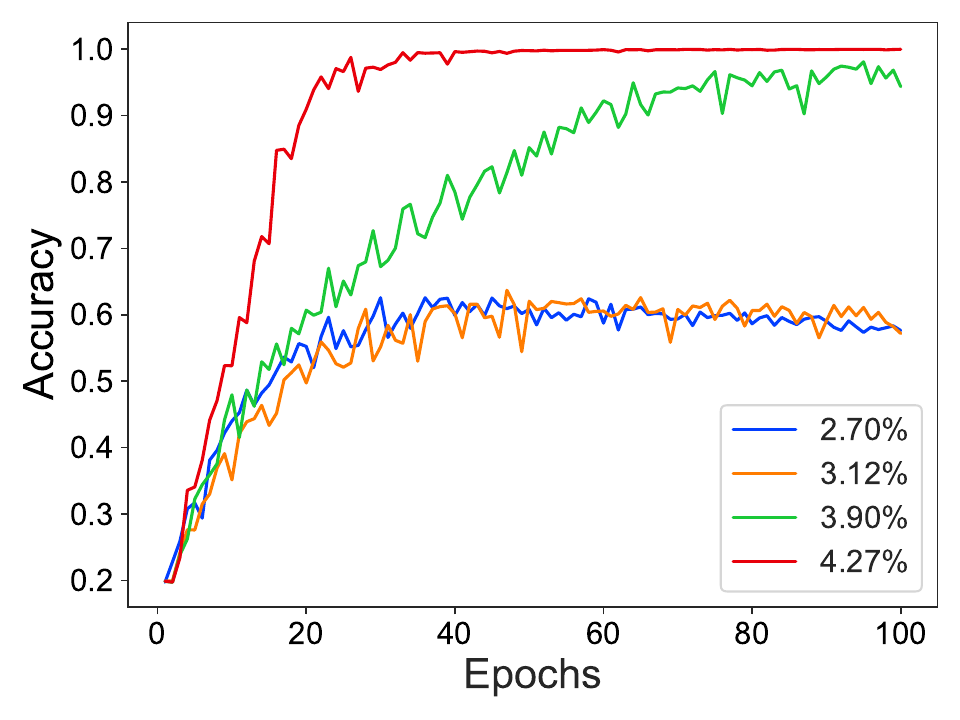}
	\caption{
		Learning curves for sequence length $16$ with different percentages
	}
	\label{fig:L16-fraction}
\end{figure}

In the Figure \ref{fig:L16-fraction} we depict the learning curves for different training set proportions for sequences of length 16 with the fixed hyperparameters, i.e., dimension $128$, head $8$, and layer $4$. 
Notably, the training set comprising $4.27\%$ of the total data was sufficient to yield an accuracy exceeding $99\%$ and 
the $3.90\%$ dataset can also achieve $90\%$ accuracy. However,
the training set with  proportion of $2.7\%$ and $3.12\%$ only achieve about $50\%$ accuracy.


\begin{figure}[H]
	\centering
	\includegraphics[scale=0.5]{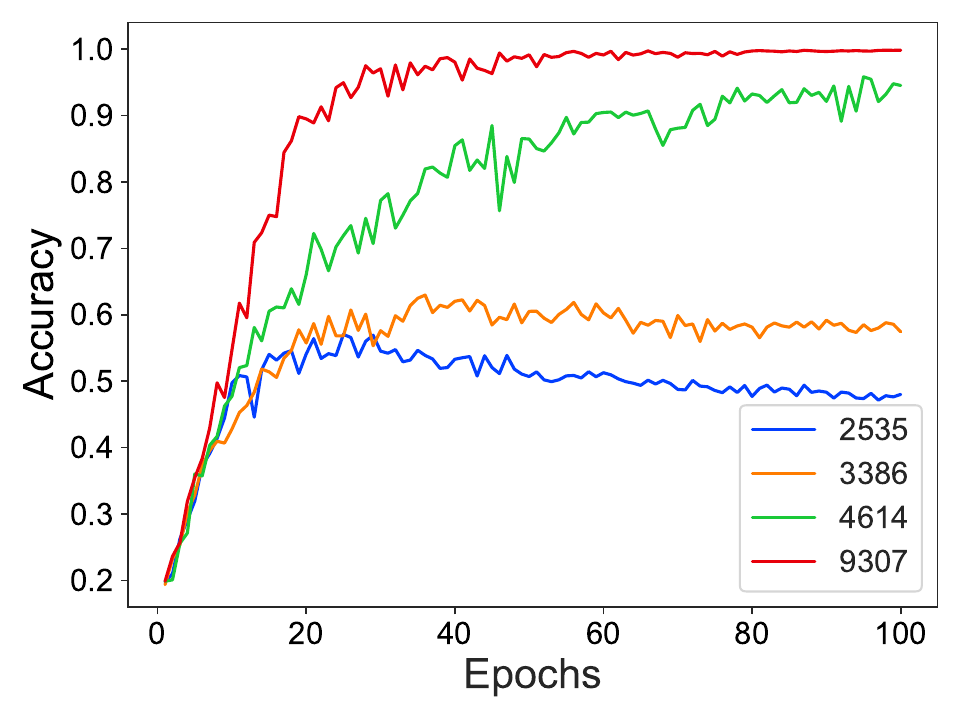}
	\caption{Model accuracy after incorporating varying number  of length-14 samples into the length-16 training set.
	}
	\label{fig:mix-16-14}
\end{figure}

Having identified the threshold of training data\footnote{Note that the threshold here is different with that in last section where the training sets are Wilson loops with length less or equal to 12.}, we add some samples with length $14$ to the training set with a proportion of 
$3.12\%$ of length $16$, while keeping the same validation and test sets as 
indicated in the Table \ref{tab:sample-number-l16}. In the Figure~\ref{fig:mix-16-14}, we show the learning curves for varying numbers of additional samples with length $14$.
These additional samples were randomly selected from each equivalence class of length 14 with the same proportion. One can see that when the number of additional length-14 samples exceeded $4,614$, the model's accuracy surpassed $95\%$. 
This result demonstrates the model's ability to generalize from sequences of length $14$ to those of length $16$.

\section{Experiments on disjoint datasets}

In previous sections, the training, validation, and test sets were sampled from each equivalence class with the same proportion. A natural concern is whether the model is simply memorizing symmetry-transformed variants of loops it has already seen, rather than learning a genuine mapping from shape to expectation value. To address this, we perform two complementary experiments using disjoint dataset splits. In both experiments below, the validation set was taken to be the same as the test set.

\subsection{Split by equivalence class}
\label{sec:disjoint_eq}

In this experiment, all Wilson loops belonging to the same equivalence class (related by lattice symmetries, cyclic permutations, or conjugation) were assigned entirely to either the training set or the test set, with no overlap between them. We trained separate models on two dataset configurations: (i) all Wilson loops with lengths $\leq 14$, and (ii) only those with length exactly 16. In each case, approximately 10\% of the equivalence classes were reserved for the test set, and the remaining data were allocated to the training set. The detailed dataset sizes are listed in Table~\ref{tab:data_stats}.

\begin{table}[htbp]
	\centering
	\begin{tabular}{lcc}
		\toprule
		Condition & Training Classes & Test Classes \\
		\midrule
		$L \leq 14$ & 534 & 50 \\
		$L = 16$   & 2897 & 320 \\
		\bottomrule
	\end{tabular}
	\caption{Number of equivalence classes in the training and test sets for the equivalence-class-disjoint experiment.}
	\label{tab:data_stats}
\end{table}

\begin{figure}[t]
	\centering
	\setlength{\tabcolsep}{-3pt}
	\subcaptionbox{Sequence lengths of 14 or
		less\label{fig:dis-b14}}{
		\includegraphics[width=0.4\textwidth]{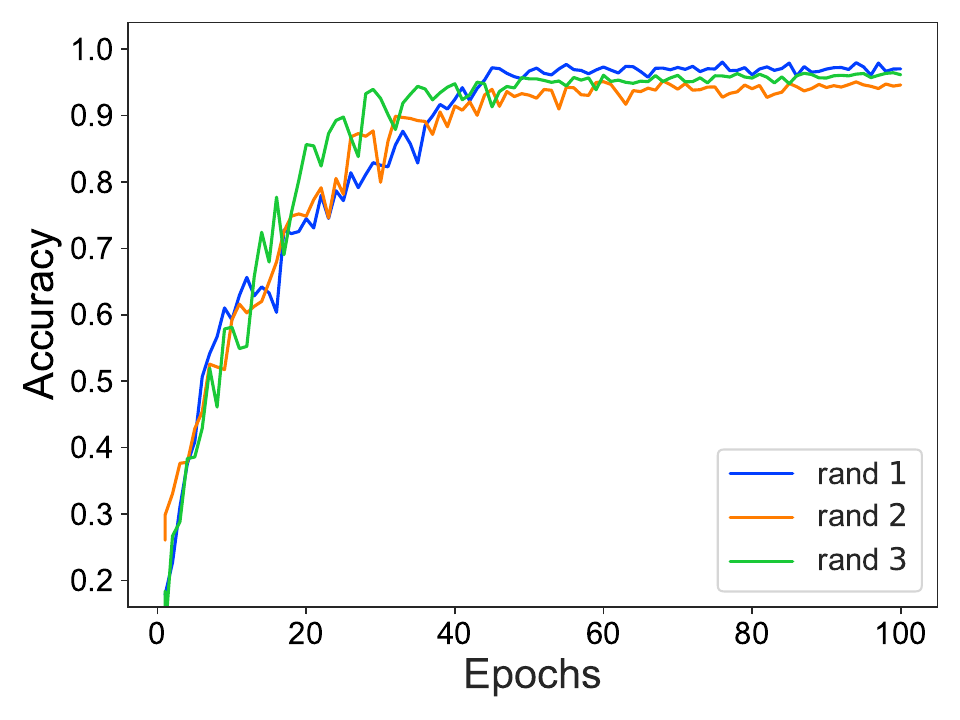}
	}
	\subcaptionbox{Sequence lengths of 16\label{fig:dis-f16}}{
		\includegraphics[width=0.4\textwidth]{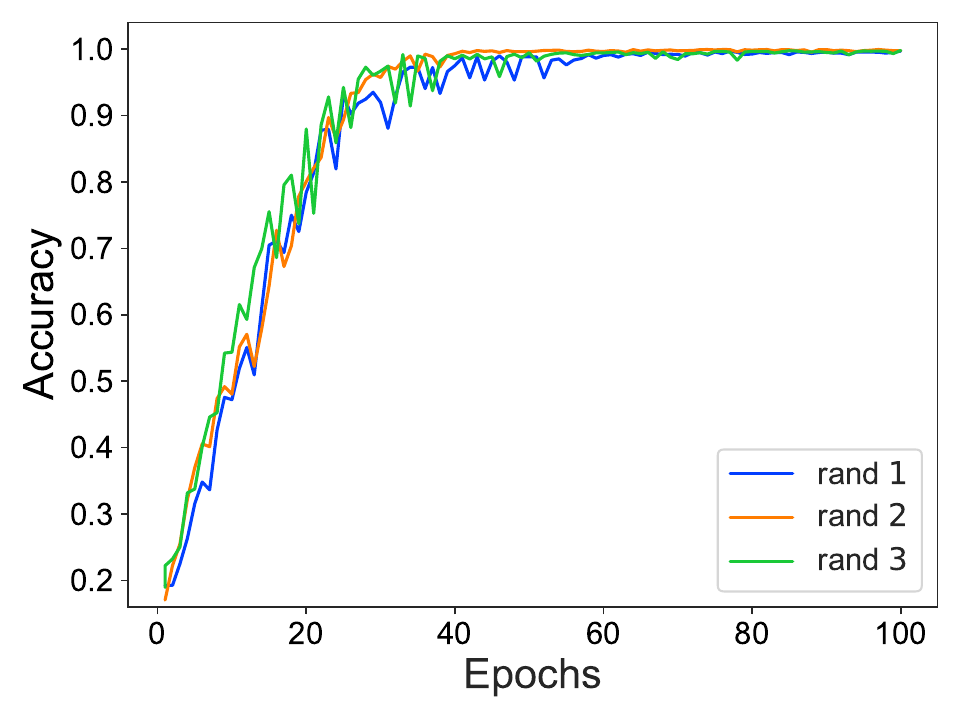}
	}
	\caption{Learning curves for the equivalence-class-disjoint experiments, showing three independent runs with different random selections of test equivalence classes. In all cases the model achieves high accuracy even though the test set consists of entirely unseen equivalence classes.}
	\label{fig:disjoint-data}
\end{figure}

To ensure the robustness of these results, we repeated each experiment three times, each time randomly selecting a different set of equivalence classes for the test set while maintaining the same proportions. The learning curves for all three runs are shown in Figure~\ref{fig:disjoint-data}. In both the $L \leq 14$ and $L = 16$ configurations, all three runs consistently achieved high prediction accuracy on their respective unseen test sets, demonstrating that the result is not sensitive to the particular choice of test equivalence classes.

These results indicate that the Transformer is capable of learning the physical mapping between the geometric shape of a Wilson loop and its expectation value, rather than simply overfitting to particular representatives of each equivalence class. To a certain extent, this suggests that the model has captured some of the genuine intrinsic features of 2D lattice Yang-Mills theory.

\subsection{Split by analytic solution}
\label{sec:disjoint_ana}

To further probe the nature of the model's learned representation, we performed a more stringent experiment in which the training and test sets were split according to distinct analytic solutions. Specifically, Wilson loops sharing the same polynomial form for their expectation value were grouped together, and the split was made such that the test set contains only analytic forms that are entirely absent from the training set. The dataset statistics for this split are shown in Table~\ref{tab:stats}. For $L \leq 14$, the training set covers 21 distinct analytic solutions while the test set contains 5 unseen ones: $u^6$, $u^{12}$, $-u^2 (u^2 + 2 u \lambda-2)$, $-u^4 ( u^2 + 4 u \lambda-2)$, and $-((u - 2 \lambda) (2 u \lambda-1))$. For $L = 16$, the training set covers 38 forms and the test set contains 5 unseen ones: $u^7$, $u^{10}$, $-u^3 ( u^2-2)$, $u^2 ( 2 u \lambda-1 )^2$, and $-u^6 ( u^2 + 4 u \lambda-2)$. Note that some building blocks like $u^2$ and $2u \lambda-1$ appear in training sets, but the combination of them do not.

\begin{table}[htbp]
	\centering
	\begin{tabular}{lccc}
		\toprule
		Condition & Dataset &  Distinct Analytical Solutions &  Loops \\
		\midrule
		$L \leq 14$ & Training & 21 & 97,232 \\
		& Test & 5 & 11,424 \\
		\cmidrule(lr){1-4}
		$L = 16$ & Training & 38 & 679,848 \\
		& Test & 5 & 82,256 \\
		\bottomrule
	\end{tabular}
	\caption{Dataset statistics for the analytic-solution-disjoint experiment. The training and test sets contain entirely different polynomial forms.}
	\label{tab:stats}
\end{table}

\begin{figure}[h]
	\centering
	\setlength{\tabcolsep}{-3pt}
	\subcaptionbox{Sequence lengths of 14 or less\label{fig:dis-ana-b14}}{
		\includegraphics[width=0.4\textwidth]{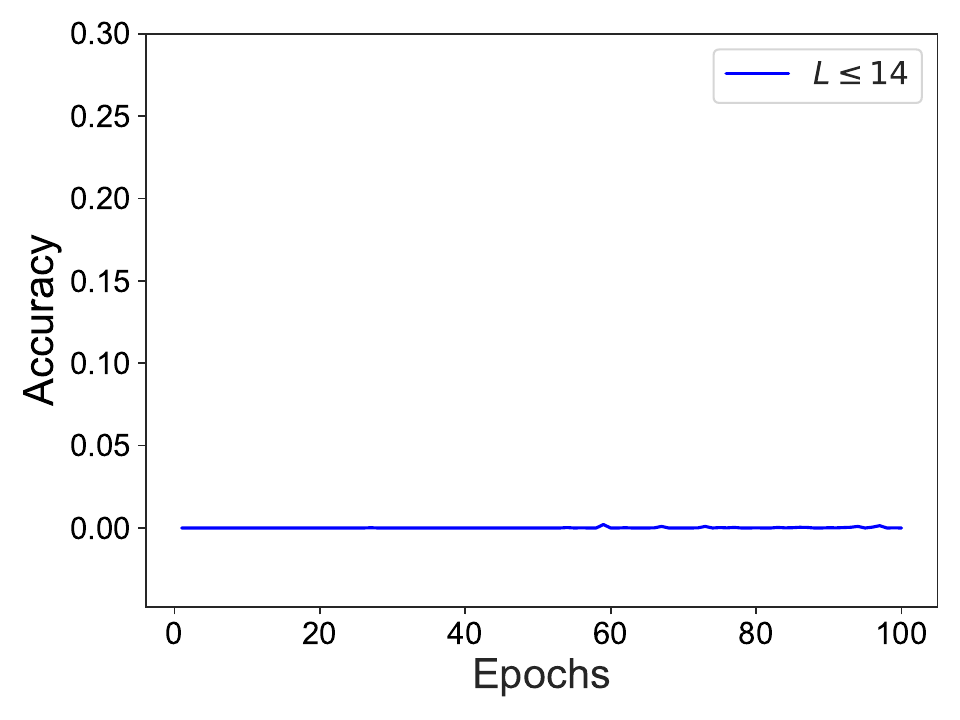}
	}
	\subcaptionbox{Sequence lengths of 16\label{fig:dis-ana-f16}}{
		\includegraphics[width=0.4\textwidth]{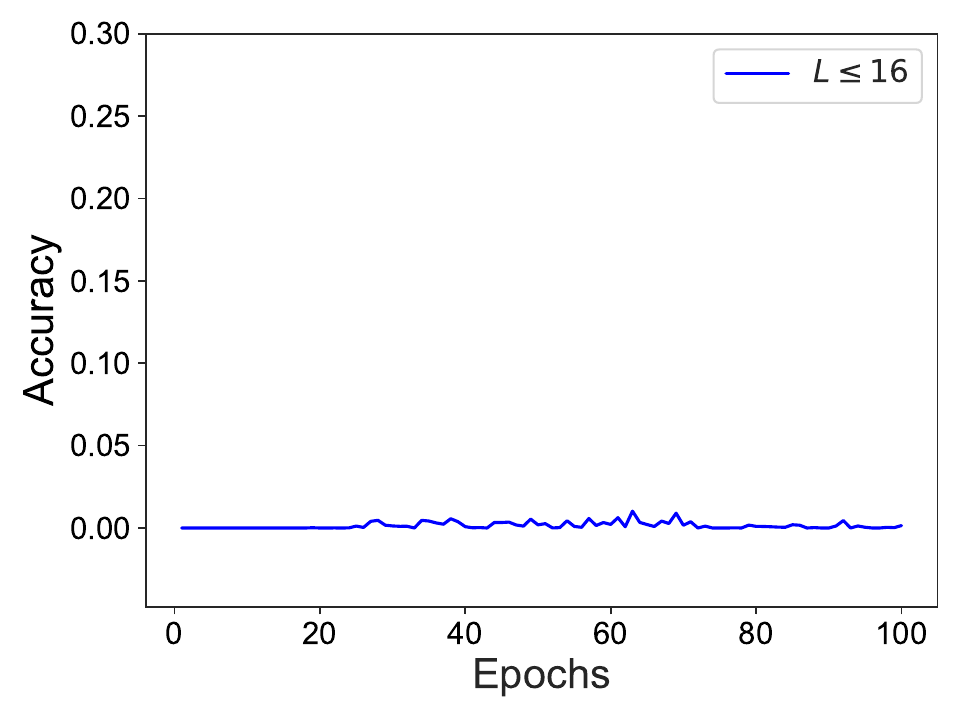}
	}
	\caption{Learning curves for the analytic-solution-disjoint experiments. The test set consists of Wilson loops whose analytic forms are entirely absent from the training set. In both cases, the model fails to generalize, with accuracy remaining below $1\%$ throughout training.}
	\label{fig:disjoint-ana}
\end{figure}

The learning curves for this experiment are shown in Figure~\ref{fig:disjoint-ana}. In stark contrast to the equivalence-class-disjoint experiment, the model fails completely in this setting: the accuracy on the test set remains below $1\%$ throughout training for both $L \leq 14$ and $L = 16$. This result demonstrates that while the Transformer can generalize across different geometric representatives of known analytic structures, it cannot extrapolate to unseen polynomial forms. In other words, the model has effectively learned to classify Wilson loops into the analytic templates present in its training data, but has not acquired a rule that would allow it to predict new functional forms.

\subsection{Extrapolation to length 18}
\label{sec:extrap_l18}

The conclusion of the previous subsection also explains the model's failure when extrapolating to longer sequences. We used the model trained on the length-16 dataset (corresponding to Figure~\ref{fig:disjoint-data}(b)) to predict the expectation values of all length-18 Wilson loops, whose statistics are summarized in Table~\ref{tab:l18}.

\begin{table}[htbp]
	\centering
	\begin{tabular}{lcc}
		\toprule
		Condition & Equivalent Classes &  Loops \\
		\midrule
		$L = 18$ & 21,762 & 6,176,976 \\
		\bottomrule
	\end{tabular}
	\caption{Statistics for $L = 18$.}
	\label{tab:l18}
\end{table}

The model achieved an overall accuracy of $19.22\%$ on the length-18 dataset. However, this aggregate figure conflates two qualitatively different regimes. Among the 6,115,320 length-18 loops whose analytic forms already appear in the $L \leq 16$ training data, the model achieves non-trivial accuracy on some forms, with per-form accuracies ranging from near zero to above $60\%$ depending on the specific polynomial form. For example, simpler forms such as $u^2$ achieve $68.43\%$ accuracy and $u^3$ achieves $69.37\%$ accuracy, while more complex forms show substantially lower performance. We show a particular example of $L=18 $ Wilson loop with $u^2$ analytic solution in Figure~\ref{fig:wilson_loop_length_18_example}. Detailed breakdowns are provided in Appendix~\ref{app:l18_accuracy}. In contrast, for the 60,192 loops with entirely new analytic forms that do not appear at length 16 or below, the model's accuracy is essentially zero.

\begin{figure}[t]
	\begin{center}
		\includegraphics[width=0.5\textwidth]{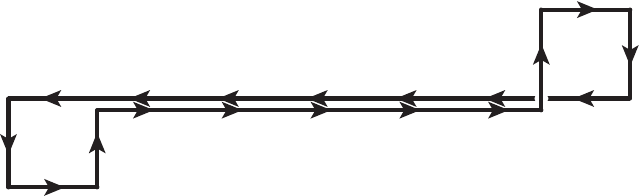}
	\end{center}
	\caption{Example of L=18 Wilson loop with $u^2$ analytic solution}\label{fig:wilson_loop_length_18_example}
\end{figure}

This result further supports the characterization of the learning task as a structured mapping over a finite catalog of outputs, and highlights the need for more advanced architectures or training strategies to achieve genuine physical generalization.

\section{Discussion}
\label{sec:discussion}

In this paper, we have used the Transformer to study the Wilson loops in 2D pure Yang-Mills lattice theory. Specifically,
we map the shape information of Wilson loop into its vacuum expectation value and we found that the Transformer achieved high accuracy within the training distribution.

Before summarizing the detailed findings, we emphasize an important caveat. The 2D lattice Yang-Mills theory studied here is exactly solvable, and as we show below, all the analytic expressions for Wilson loops with length up to 16 belong to only 46 distinct polynomial forms built from 10 basic factors. This means the output space of the learning task is finite and restricted. Nevertheless, as demonstrated by the disjoint equivalence class experiment in section~\ref{sec:disjoint_eq}, the model does exhibit meaningful generalization: it can correctly predict expectation values for geometrically distinct Wilson loops that it has never seen during training, as long as their analytic forms are among those already learned. The model's limitation lies in its inability to extrapolate to entirely new polynomial structures, as shown in sections~\ref{sec:disjoint_ana} and \ref{sec:extrap_l18}.


First, we study the impact of choices of hyperparameters of the models using input sequences of fixed length $14$. 
By various combinations of three hyperparameters, i.e., the model dimensions, attention heads, and layers, we see that the choice should be balanced among them to have a robust, good learning performance. For Wilson loops with length 14, the choice
with dimension $\dm = 128$,
layer count $\nl = 4$,
and attention heads $\nh = 8$ seems to be a good one. 

Secondly, we investigate the impact of training set proportions on model learning efficiency with the datasets comprised by sequences of length $12$ or less. 
We discover a threshold for proportion of training set, when the proportion falls below this threshold, the model fails to learn the task. We also find that when the proportion approaches this threshold, the learning curve exhibits fluctuations depending on the initial conditions.


We also conducted a mixed length learning. Specifically, we add some data with length $14$ to the training data with length $16$ and we find that this can indeed improve the accuracy for data of length 16.
This experiment suggests that the model can benefit from low-length data and assist in the learning of high-length data, which may hold significance for training with large volumes of high-length data.


One observation is that, even with a very small proportion of the total dataset to do the training, as seen in the Table \ref{tab:sample-number-l16}, we are able to learn the task. This can be understood from the limited complexity of the output space, as discussed in Section~\ref{sec:conventions_and_setup}: all analytic expressions for Wilson loops with length $L \leq 16$ belong to only $46$ distinct polynomial forms built from $10$ basic factors. This limited variability reduces the complexity of the learning task, enabling the model to generalize effectively from the small training data.

An important aspect of this study is understanding the model's generalization capability. As shown in subsections~\ref{sec:disjoint_ana} and \ref{sec:extrap_l18}, the Transformer can generalize across different geometric representatives of known analytic structures, but cannot extrapolate to unseen polynomial forms — whether they appear within the same sequence length or in longer sequences. The model effectively learns a finite dictionary of known structures rather than discovering new relations between geometry and physics. A key challenge for future research is developing models that can bridge the gap from supervised pattern fitting to true physical generalization. We stress that the encoding scheme for the lattice links proposed in this paper provides a useful foundation for building these next-generation models and designing more sophisticated experiments to achieve this goal.


Finally, for a given Wilson loop, there are a few important characteristics for the description of its geometry, i.e., the area, the number of crossing lines, and the way of crossing. It is plausible that the analytic expression could be determined by these geometric invariants. Establishing such a connection rigorously would be an interesting direction for future studies.  


\section*{Acknowledgments}
Zeyu Li, Guorui Zhu and Gang Yang are supported by the Chinese Academy of Sciences (Grant No. YSBR-101) and by the National Natural Science Foundation of China (Grants No. 12425504, 12175291, 12047503, 12247103).
Ming-xing Luo is supported by National Natural Science Foundation of China under contract No. U2230402 and NSFC-CERN W2443010. 
Bo Feng is supported by the National Natural
Science Foundation of China (NSFC) through Grants No.12535003, No.11935013, No.11947301, No.12047502. 
Jiaqi Chen is supported by Science Foundation of China University of Petroleum,
Beijing (No.2462025YJRC019). 

\appendix
\section{Impact of Dataset Randomness on Learning Trajectories}
\label{sec:dataset-randomness}

In the section \ref{sec:fractrain}, we have seen the impact of the randomness of the initial condition on the learning
performance. In this section, we will show the effect of the randomness of training sets on the learning ability. 

\begin{figure}[t]
	\centering
	\includegraphics[width=0.5\textwidth]{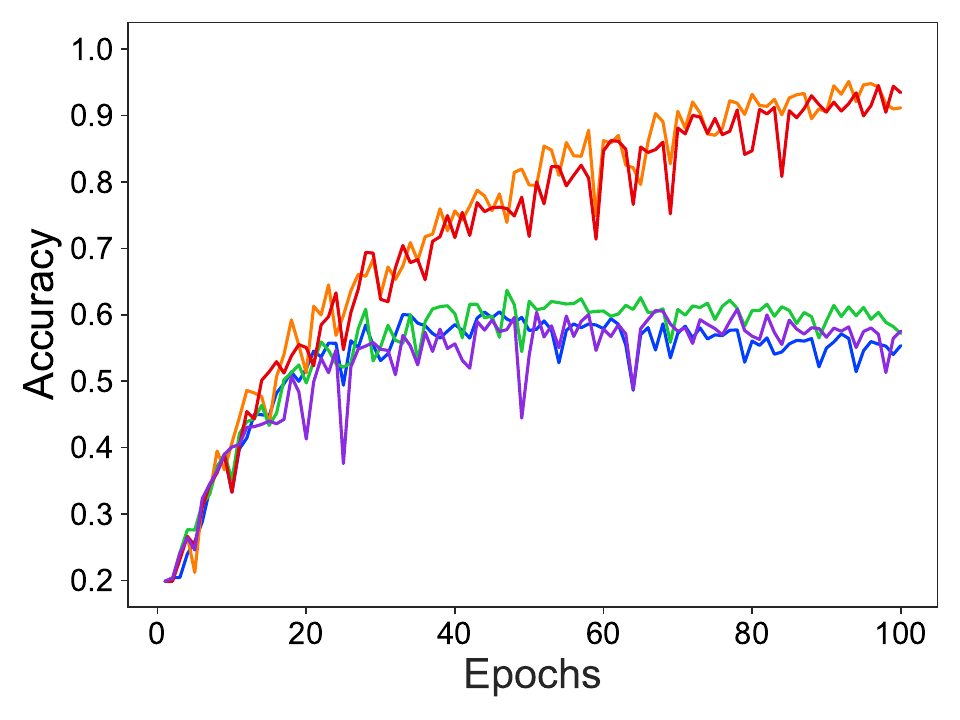}
	\caption{
		Learning curves trained by different random datasets. All the training sets contain 23,815 data samples.
	}
	\label{fig:repeat-3p5}
\end{figure}

In the section  \ref{sec:mixtrain}, we found that
training data with the proportions of $2.7\%$ and $3.12\%$ fail to learn the Wilson loops with length 16. Now we redo this example of proportion  $3.12\%$, i.e., training samples comprising 23,815 training samples. 
In the Figure \ref{fig:repeat-3p5} we show the learning curves of models trained on different randomly sampled length-16 sequence datasets, exhibiting a separated banded structure. The figure demonstrates that not all training datasets fail to learn the task. In five random tests, two training sets successfully increased accuracy to over $90\%$. This phenomenon shows that if the dataset has a good representation of the whole dataset, even if it contains small proportion of samples, it still can perform well. But in general, we should not expect a small proportion to have a good representation due to randomness.

\begin{figure}[t]
	\centering
	\subcaptionbox{
		Learning curves trained by different random datasets. All the training sets contain 32,511 data samples.
		\label{fig:repeat-L16-32511}
	}{
		\includegraphics[width=0.35\textwidth]{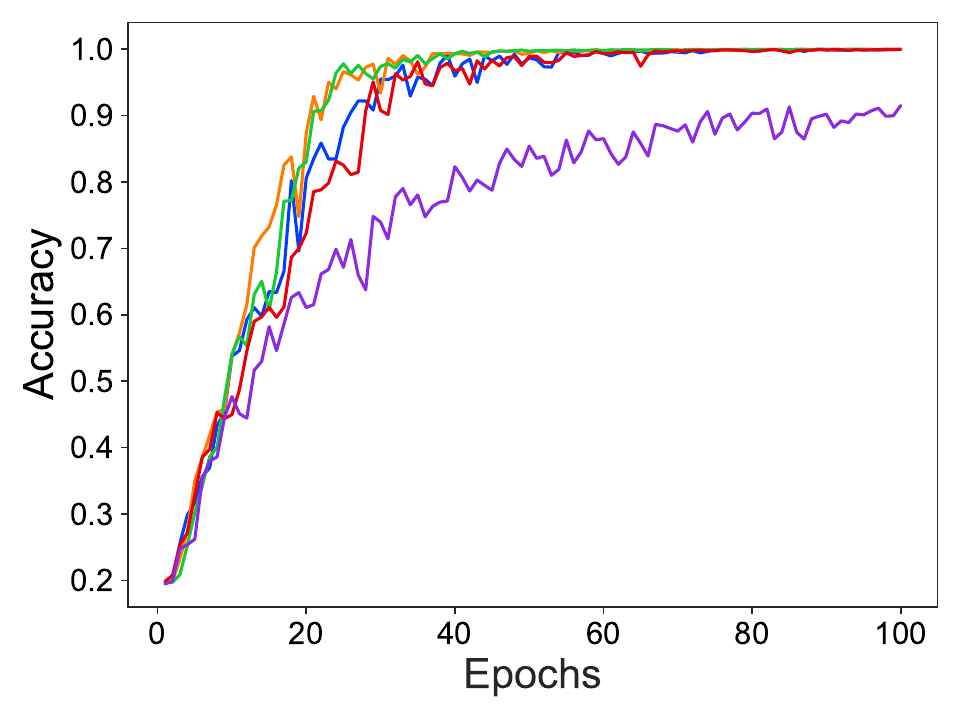}
	}~~~
	\subcaptionbox{
		Learning curves trained by different random datasets. 
		All the training sets contain 23,815 length-16 data samples and 18,423 length-14 data samples.
		\label{fig:repeat-add-18423}
	}{
		\includegraphics[width=0.35\textwidth]{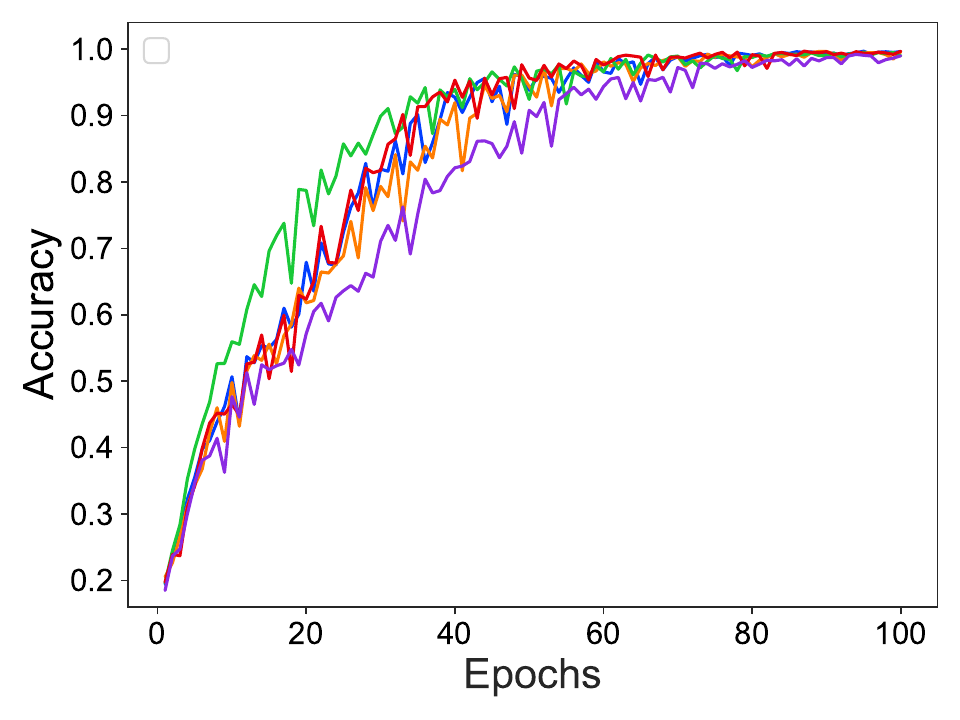}
	}
	\caption{Learning curves with training sets which is larger than threshold range.}
	\label{fig:repeat-large}
\end{figure}

Next, we increase the proportion of training data to reveal the effect of randomness. In the Figure \ref{fig:repeat-L16-32511}, $4.27\%$ of the length-16 dataset has been used. With all five 
training sets, they do have a good performance, although one of them is a little bit worse than the others. Similarly, 
for the mixed training data with length-14 and length-16 sequences in Figure \ref{fig:repeat-add-18423}, all five 
training sets perform well.

Experiments in this section suggest that increased dataset size mitigates variance introduced by dataset randomness, leading to more consistent convergence patterns. In other words, to eliminate the impact of randomness 
from initialization, the training data, or other places, a sufficient number of samples in the training dataset
should be required. 

\begin{figure}[t]
	\centering
	\includegraphics[width=0.5\textwidth]{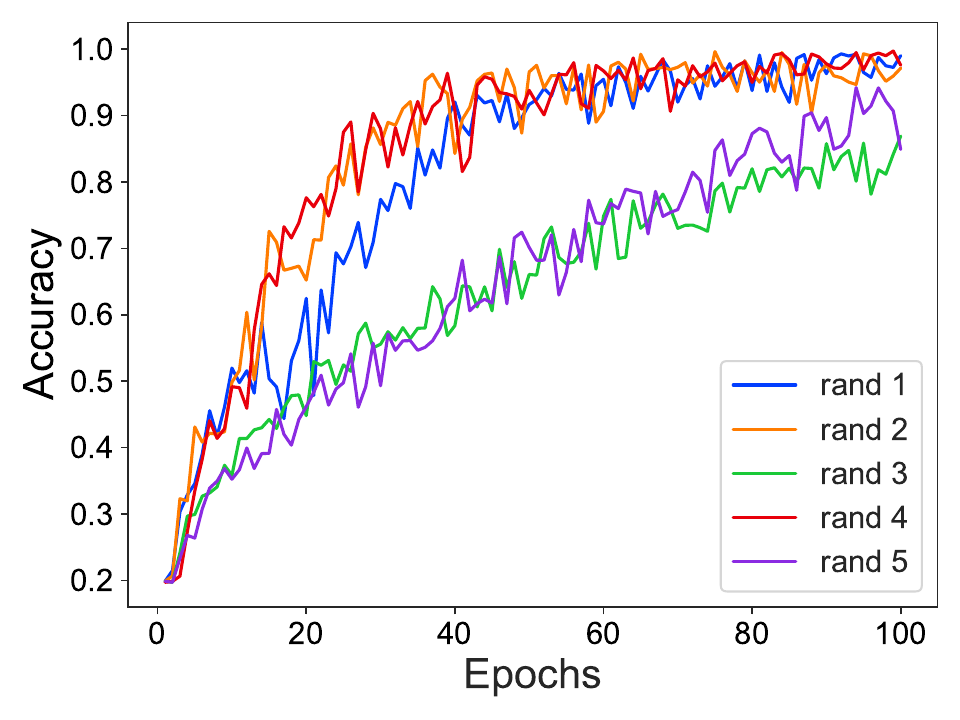}
	\caption{Accuracy fluctuations across different training sets.}
	\label{fig:L16-repeat-64}
\end{figure}

Before ending this section, we want to mention that although a larger size of training datasets can reduce the impact of randomness, the choice of hyperparameters should be balanced. As shown in the Figure \ref{fig:L16-repeat-64}, with the
choice of dimension $64$, head $4$, and layer $4$, although the average learning accuracy is increasing over epochs, the fluctuations remain significant even after approximately 60 epochs.

\section{Per-form accuracy for length-18 extrapolation}
\label{app:l18_accuracy}

Table~\ref{tab:l18_accuracy} provides the detailed prediction accuracy for each analytic form when the model trained on $L \leq 16$ data is applied to length-18 Wilson loops. The table lists each polynomial form, the number of length-18 loops with that form, and the corresponding prediction accuracy.

Forms appearing in the training data (i.e., present in $L \leq 16$) show non-zero accuracy, with performance varying significantly across different forms. Forms that do not appear in the training data (marked with 0 accuracy) represent entirely new polynomial structures that emerge only at length 18.

\begin{table}[htbp]
	\centering
	\footnotesize
	\setlength{\tabcolsep}{4pt}
	\renewcommand{\arraystretch}{0.85}
	\begin{tabular}{lcc}
		\toprule
		Analytic Form & Count & Accuracy \\
		\midrule
		$u^{17}$ & 3,168 & 0 \\
		$u^{18}$ & 1,080 & 0 \\
		$u^{19}$ & 288 & 0 \\
		$u^{20}$ & 72 & 0 \\
		$u^2 + u^4 - u^6$ & 432 & 0 \\
		$u^3 + u^5 - u^7$ & 288 & 0 \\
		$2u^5 - u^7$ & 3,456 & 0 \\
		$2u^6 - u^8$ & 792 & 0 \\
		$u - 2u^2\lambda + 2u^4\lambda$ & 288 & 0 \\
		$3u^4 - 2u^6 - 4u^5\lambda$ & 432 & 0 \\
		$u^2 + u^4 - u^6 - 2u^5\lambda$ & 432 & 0 \\
		$3u^5 - 2u^7 - 4u^6\lambda$ & 288 & 0 \\
		$u^3 + u^5 - u^7 - 2u^6\lambda$ & 288 & 0 \\
		$2u^5 - u^7 - 2u^6\lambda$ & 6,912 & 0.00376 \\
		$3u^6 - 2u^8 - 6u^7\lambda$ & 864 & 0 \\
		$2u^6 - u^8 - 2u^7\lambda$ & 1,584 & 0.01326 \\
		$3u^7 - 2u^9 - 6u^8\lambda$ & 864 & 0 \\
		$2u^7 - u^9 - 4u^8\lambda$ & 9,504 & 0 \\
		$4u^8 - 3u^{10} - 8u^9\lambda$ & 144 & 0 \\
		$2u^8 - u^{10} - 4u^9\lambda$ & 2,592 & 0 \\
		$2u^9 - u^{11} - 4u^{10}\lambda$ & 432 & 0 \\
		$u^9 - 2u^{10}\lambda$ & 10,800 & 0 \\
		$u^{10} - 2u^{11}\lambda$ & 2,880 & 0 \\
		$u^{11} - 2u^{12}\lambda$ & 720 & 0 \\
		$u - 4u^2\lambda + 2u^4\lambda + 4u^3\lambda^2$ & 288 & 0 \\
		$u^2 - 6u^3\lambda + 8u^4\lambda^2$ & 2,160 & 0 \\
		$2u^2 - u^4 - 8u^3\lambda + 2u^5\lambda + 8u^4\lambda^2$ & 432 & 0 \\
		$u^3 - 4u^4\lambda + 4u^5\lambda^2$ & 4,032 & 0 \\
		$u^5 - 2u^4\lambda - 2u^6\lambda + 4u^5\lambda^2$ & 2,016 & 0 \\
		$u^3 - 6u^4\lambda + 8u^5\lambda^2$ & 576 & 0 \\
		$2u^3 - u^5 - 8u^4\lambda + 2u^6\lambda + 8u^5\lambda^2$ & 288 & 0 \\
		$u^4 - 4u^5\lambda + 4u^6\lambda^2$ & 1,080 & 0 \\
		$u^6 - 2u^5\lambda - 2u^7\lambda + 4u^6\lambda^2$ & 432 & 0 \\
		$u - 2\lambda - 4u^2\lambda + 8u\lambda^2 + 4u^3\lambda^2 - 8u^2\lambda^3$ & 144 & 0 \\
		$u - 8u^2\lambda + 4u\lambda^2 + 12u^3\lambda^2 - 8u^2\lambda^3$ & 144 & 0 \\
		\bottomrule
	\end{tabular}
	\caption{Prediction accuracy for each analytic form on length-18 Wilson loops. Forms with 0 accuracy are new polynomial structures that do not appear in the $L \leq 16$ training data.}
	\label{tab:l18_accuracy}
\end{table}

For comparison, Table~\ref{tab:l16_accuracy} in the next page shows the per-form accuracy on the length-16 training data itself, demonstrating the model's performance on forms it has seen during training.

\begin{table}[htbp]
	\centering
	\footnotesize
	\setlength{\tabcolsep}{4pt}
	\renewcommand{\arraystretch}{0.85}
	\begin{tabular}{lcc}
		\toprule
		Analytic Form & Count & Accuracy \\
		\midrule
		$u^2$ & 45,792 & 0.6843 \\
		$u^3$ & 328,896 & 0.6937 \\
		$u^4$ & 843,264 & 0.4462 \\
		$u^5$ & 1,108,944 & 0.2011 \\
		$u^6$ & 941,688 & 0.1327 \\
		$u^7$ & 650,880 & 0.0914 \\
		$u^8$ & 428,472 & 0.0466 \\
		$u^9$ & 281,232 & 0.0180 \\
		$u^{10}$ & 183,600 & 0.0078 \\
		$u^{11}$ & 120,096 & 0.0077 \\
		$u^{12}$ & 81,720 & 0.0000 \\
		$u^{13}$ & 51,696 & 0.0000 \\
		$u^{14}$ & 31,104 & 0 \\
		$u^{15}$ & 16,416 & 0 \\
		$u^{16}$ & 7,848 & 0 \\
		$2u^3 - u^5$ & 3,600 & 0.0869 \\
		$2u^4 - u^6$ & 8,208 & 0.0127 \\
		$u - 2u^2\lambda$ & 9,792 & 0.5942 \\
		$u^2 - 2u^3\lambda$ & 93,312 & 0.4211 \\
		$2u^3 - u^5 - 4u^4\lambda$ & 4,896 & 0.2063 \\
		$u^3 - 2u^4\lambda$ & 222,624 & 0.2313 \\
		$2u^3 - u^5 - 2u^4\lambda$ & 7,200 & 0.1393 \\
		$2u^4 - u^6 - 4u^5\lambda$ & 16,560 & 0.0121 \\
		$u^4 - 2u^5\lambda$ & 226,944 & 0.0439 \\
		$2u^4 - u^6 - 2u^5\lambda$ & 16,416 & 0.0177 \\
		$3u^5 - 2u^7 - 6u^6\lambda$ & 288 & 0 \\
		$2u^5 - u^7 - 4u^6\lambda$ & 23,904 & 0.0031 \\
		$u^5 - 2u^6\lambda$ & 147,744 & 0.0285 \\
		$2u^6 - u^8 - 4u^7\lambda$ & 18,288 & 0.0005 \\
		$u^6 - 2u^7\lambda$ & 84,960 & 0.0153 \\
		$u^7 - 2u^8\lambda$ & 51,984 & 0.0003 \\
		$u^8 - 2u^9\lambda$ & 26,496 & 0.0000 \\
		$1 - 4u\lambda + 4u^2\lambda^2$ & 576 & 0.1528 \\
		$u^2 - 2u\lambda - 2u^3\lambda + 4u^2\lambda^2$ & 1,152 & 0.0885 \\
		$u - 4u^2\lambda + 4u^3\lambda^2$ & 6,480 & 0.1526 \\
		$u^3 - 2u^2\lambda - 2u^4\lambda + 4u^3\lambda^2$ & 6,192 & 0.0896 \\
		$u^2 - 4u^3\lambda + 4u^4\lambda^2$ & 9,720 & 0.0003 \\
		$u^4 - 2u^3\lambda - 2u^5\lambda + 4u^4\lambda^2$ & 6,336 & 0 \\
		\bottomrule
	\end{tabular}
	\caption{Prediction accuracy for each analytic form on length-16 Wilson loops (training data). This shows the model's performance on forms it has encountered during training.}
	\label{tab:l16_accuracy}
\end{table}

	
\end{document}